\documentclass[%
 reprint,
nofootinbib,https://www.overleaf.com/project/6151c250894695bd52e675b8
 amsmath,amssymb,
 aps,
floatfix,
longbibliography,
]{revtex4-1}

\usepackage[english]{babel}
\usepackage[utf8]{inputenc}
\usepackage{graphicx}% Include figure files
\usepackage{graphics}% Include figure files
\usepackage{dcolumn}% Align table columns on decimal point
\usepackage{bm}% bold math
\usepackage[mathlines]{lineno}% Enable numbering of text and display math
\usepackage{xcolor}
\usepackage{tikz}
\DeclareUnicodeCharacter{2212}{-}
\usepackage{pict2e} % for drawing polygons
\usepackage[permil]{overpic}
\usepackage{multirow}
\usepackage{threeparttable}
\usepackage{supertabular}
\usepackage{footnote}
\usepackage{tablefootnote}
\usepackage{longtable} 
\usepackage{adjustbox}
\usepackage{placeins}
\begin{document}
\preprint{LHCb-implications-1}

\title{Energy deposition studies in the LHCb insertion region:\\from the validation to a step into the Hilumi challenge}% Force line breaks with \\

\author{Alessia Ciccotelli}
 \email{alessia.ciccotelli@cern.ch}
\affiliation{University of Manchester, Manchester, United Kingdom.
}%
\affiliation{CERN, CH 1211 Geneva 23, Switzerland.}%Lines break automatically or can be forced with \\

\author{Robert B. Appleby}
\affiliation{University of Manchester, Manchester, United Kingdom.
}%
\author{Francesco Cerutti, Kacper Bilko, Luigi Salvatore Esposito, Ruben Garcia Alia, Anton Lechner, Andrea Tsinganis}%
\affiliation{%
 CERN, CH 1211 Geneva 23, Switzerland.%\\
}%


%\affiliation{
%\affiliation{%


\date{\today}% It is always \today, today,

\begin{abstract}
The LHCb (Large Hadron Collider beauty) experiment at CERN aims at achieving a significantly higher luminosity than originally planned by means of two major upgrades: the Upgrade I that took place during the Long Shutdown~2 (LS2) and the Upgrade II foreseen for LS4. 
Such an increase in instantaneous and integrated luminosity with respect to the design values requires to reassess the radiation exposure of LHC magnets, cryogenics and electronic equipment placed in the Insertion Region 8 (IR8) around LHCb. 
Monte Carlo simulations are a powerful tool to understand and predict the interaction between particle showers and accelerator elements, especially in case of future scenarios. For this purpose, their validation through the comparison with available measurements is a relevant step. A detailed IR8 model, including the LHCb detector, has been implemented with the FLUKA code. The objective of this study is to evaluate radiation levels due to proton--proton collisions and benchmark the predicted dose values against Beam Loss Monitor (BLM) measurements performed in 2018. Finally, we comment on the upcoming LHC run (Run~3), featuring a first luminosity jump in LHCb.

\end{abstract}

%Use showkeys class option if keyword display desired
\maketitle

%\tableofcontents

%\section{\label{sec:level1}First-level heading:\protect\\ The line
%break was forced \lowercase{via} \textbackslash\textbackslash}
\section{Introduction}

During Run~2 (2015-2018), the Large Hadron Collider (LHC) at CERN collided $6.5~\mathrm{TeV}$ proton beams, achieving more than twice its design instantaneous luminosity in the ATLAS and CMS detectors.
The LHCb experiment, located in the Insertion Region 8 (IR8), was designed to work at a lower luminosity than ATLAS and CMS, as shown in Table~\ref{tab:table3}, implying a lower need for protection of the LHC elements from the collision debris and therefore a different layout around the Interaction Point (IP).
During the Long Shutdown~2 (LS2), the LHCb detector has been upgraded in order to increase its statistical precision, undergoing its so-called Upgrade I~\cite{Bediaga:1443882}. Previously, LHCb obtained a set of impressive physics results such as the first observation of CP violation in $B^0_s$ and charged $B$ meson decays. Even if no significant signs of new physics have been found, the data analysis showed the emergence of a disagreement with the Standard Model predictions in the measurement of rare $B$ decays at the level of 2-3 sigma~\cite{LHCb:2020lmf,LHCb:2021trn,LHCb-CONF-2020-002}. The increased luminosity of Run~3 (2022-2025) and the following Run~4, as reported in Table~\ref{tab:table3}, will allow LHCb to reduce the uncertainty of these measurements and possibly unveiling new phenomena~\cite{Aaij:2636441}. The Upgrade~I of LHCb mainly concerns the tracking system and the electronics of most sub-detectors due to the conversion into a full software trigger system and a trigger-less readout system, de facto removing the limitation from hardware trigger technology. As a result of the Upgrade I, the LHCb experiment expects to sustain a peak luminosity of $2\cdot10^{33}\,\mathrm{cm}^{−2}\,\mathrm{s}^{−1}$ which is around 5 times higher than the maximum value reached during the Run~2 proton operation. 
\begin{table*}[!ht]%The best place to locate the table environment is directly after its first reference in text
\caption{Overview of operational conditions for proton-proton collisions~\cite{LHCb_Luminosity,CERN_ATLAS_Luminosity,CERN_CMS_Luminosity,CERN_LHC_schedule,HLLHC_Luminosity,HLLHC-lumirampup}}.
\label{tab:table3}
\begin{ruledtabular}
\begin{threeparttable}
%\begin{tabular}[c]{m{1cm} m{1cm} | m{5cm} |m{5cm}|m{5cm}}
\begin{tabular}[c]{c c |c | c |c| c}
%\textrm{Left\footnote{Note a.}}&
\multicolumn{2}{c|}{\textrm{Period}} & \textrm{Run~2} &  \textrm{Run~3}\tnote{a} &\textrm{Run~4} & \textrm{Post-LS4}\\ [0.5ex]
\colrule
\multicolumn{2}{c|}{\textrm{Beam energy}} &$6.5\,\mathrm{TeV}$ &$6.8\,\mathrm{TeV}$& $6.8-7\,\mathrm{TeV}$& $7\,\mathrm{TeV}$\\ [0.5ex]
\colrule
\multirow{2}{4em}{\textrm{LHCb}}& $\mathcal{L}_{\rm int}$ & $6.6\,\mathrm{fb}^{−1}$\tnote{b} & $25-30\,\mathrm{fb}^{−1}$ & $25-30\,\mathrm{fb}^{−1}$ & $400 \,\mathrm{fb}^{−1}$ by the end of HL-LHC\tnote{d}\\[0.5ex]
& $\mathcal{L}^{\rm peak}$ & $4\cdot10^{32}\,\mathrm{cm}^{−2}\,\mathrm{s}^{−1}$ & $2\cdot10^{33}\,\mathrm{cm}^{−2}\,\mathrm{s}^{−1}$  & $2\cdot10^{33}\,\mathrm{cm}^{−2}\,\mathrm{s}^{−1}$  & $1.5\cdot10^{34}\,\mathrm{cm}^{−2}\,\mathrm{s}^{−1}$\tnote{d} \\[0.5ex]
\colrule
%$\mathcal{L}_{\rm int}$ & $147\,\mathrm{fb}^{−1}$ (ATLAS),$151\,\mathrm{fb}^{−1}$ (CMS)
\multirow{2}{4em}{\textrm{GPDs}\tnote{c} }& $\mathcal{L}_{\rm int}$ & $160\,\mathrm{fb}^{−1}$ & $250-300\,\mathrm{fb}^{−1}$ & $560\,\mathrm{fb}^{−1}$ & $4000 \,\mathrm{fb}^{−1}$ by the end of HL-LHC\\[0.5ex]
& $\mathcal{L}^{\rm peak}$ & $2\cdot10^{34}\,\mathrm{cm}^{−2}\,\mathrm{s}^{−1}$ & $2\cdot10^{34}\,\mathrm{cm}^{−2}\,\mathrm{s}^{−1}$ & $5\cdot10^{34}\,\mathrm{cm}^{−2}\,\mathrm{s}^{−1}$ & $5-7.5\cdot10^{34}\,\mathrm{cm}^{−2}\,\mathrm{s}^{−1}$ \\[0.5ex]
%https://lhcb.web.cern.ch/speakersbureau/html/PerformanceNumbers.html
%http://lhc-commissioning.web.cern.ch/performance/Run-3-targets.html
%https://twiki.cern.ch/twiki/bin/view/AtlasPublic/LuminosityPublicResultsRun2#Luminosity_Plots_for_multiple_Ru
%https://twiki.cern.ch/twiki/bin/view/CMSPublic/LumiPublicResults#LHC_and_CMS_luminosity_records
%https://edms.cern.ch/ui/file/2364638/1.0/GL_HLLHC_performance_parameters.pdf
% https://edms.cern.ch/ui/file/2364638/1.0/GL_HLLHC_performance_parameters.pdf
% https://lbgroups.cern.ch/online/OperationsPlots/index.htm
% https://indico.cern.ch/event/897816/contributions/3786241/attachments/2007136/3352475/SLIDESlogo.pdf
% https://hilumilhc.web.cern.ch/sites/default/files/HL-LHC_Janvier2022.pdf
\end{tabular}

\begin{tablenotes}
   \item[a] From 2022 to 2025.
   \item[b] The integrated luminosity previously collected during Run~1 is $3.4\,\mathrm{fb}^{−1}$.
   \item[c] General purpose detectors installed at the LHC: ATLAS and CMS. 
   \item[d] In case of the Upgrade~II of LHCb. 
  \end{tablenotes}
\end{threeparttable}
\end{ruledtabular}
\end{table*}
Moreover, the Upgrade II~\cite{LHCbCollaboration:2776420} envisaged for the LS4, profiting from the use of new detector technologies, aims to operate at $1.5\cdot10^{34}\,\mathrm{cm}^{−2}\,\mathrm{s}^{−1}$ and a center of mass energy $\sqrt{s}=14\,\mathrm{TeV}$ and to reach an integrated luminosity of $400 \,\mathrm{fb}^{−1}$ by the end of the High Luminosity LHC (HL-LHC) era. 

The challenges implied by this substantial increase in luminosity concern not only the detector but also the LHC machine, subject to a much higher collision debris power. In particular, the consequences of its impact on accelerator elements, surrounding devices and environment have to be anticipated and be under control. Because of the complexity of the LHC layout and infrastructure, radiation levels are calculated by means of Monte Carlo simulations. We used the FLUKA code~\cite{FLUKA,FLUKA2021,FLUKA:2015}, which is a multi-purpose Monte Carlo code widely employed to describe particle transport and interactions in many applications and extensively tested at CERN and in other laboratories for such complex geometries and high energy physics problems. In this regard, several authors have obtained a remarkable agreement comparing FLUKA calculations with experimental measurements as well as other particle simulation tools \cite{PhysRevAccelBeams.22.071003,Lechner:2674116,Mereghetti:2045506,Mokhov:2206746}. This way, the design of new machines and pieces of equipment can be driven by energy deposition predictions that define operation challenges and material lifetime.

In particular, depending on the level and distribution of the power density absorbed by the magnet coils as a consequence of the radiation impact, superconducting cables may warm above a critical temperature and so lose their ability to conduct electricity without resistance. Such a sudden and violent transition to the normal conducting state, implying the loss of the required magnetic field and the beam dump, is referred to as \textit{quench}. Although the superconducting state can later be restored and the beam re-injected, the collider operation at a given luminosity is not possible if the respective collision debris leads to regularly surpass the quench limit. Moreover, both for superconducting and normal conducting magnets, the coil material, especially the insulator, progressively deteriorates as a function of accumulated radiation dose, which thereby limits the equipment lifetime. In this paper, we evaluate these relevant quantities in view of the LHCb luminosity increase targeted in the upcoming LHC Run~3 (see Table~\ref{tab:table3}), in order to systematically confirm the upgrade sustainability from the accelerator point of view and anticipate possible issues.

There has been a variety of FLUKA models of IR8 developed at CERN in the last ten years for different types of calculations. In 2010, a FLUKA study evaluated dose and fluence levels in the experimental cavern for Run~1 and Run~2~\cite{1099674} to assess radiation induced effects on electronics. In addition, FLUKA simulations were used to evaluate the machine induced background to the LHCb experiment~\cite{Appleby:1238789}. Other investigations on the impact of the collision debris on the machine elements were carried out for Upgrade I scenarios based on the detector model and the optics corresponding to Run~2~\cite{LHCbUpgrade2020_Cerutti,Esposito:1636136}. From the perspective of the LHCb collaboration, radiation levels in the immediate surroundings of the detector were calculated and analyzed with FLUKA \cite{Karacson:2243499} during Run~2.
No simulation studies have been carried out on the entire cavern for the Run~3 configuration, especially where LHC electronics racks are placed. Furthermore, while a comparison of the predicted dose values with Beam Loss Monitor (BLM) measurements on the right-side magnets was already published for Run~1~\cite{PhysRevAccelBeams.22.071003}, no benchmarking was performed yet for Run~2 at a center of mass energy of $13\,\mathrm{TeV}$, nor for the opposite (asymmetric) side.
Therefore, the aforementioned models have been now merged and improved for the present study, combining a more accurate implementation of the LHCb detector with an extensive description of the LHC infrastructure and beamline, and this way producing an updated framework for future investigations. These will be especially devoted to
the implications of the Upgrade II, after the preliminary analysis~\cite{Efthymiopoulos:2319258,LHCbUpgrade2020_Cerutti} that identified the key points towards which dedicated simulations should be directed, building on the results described in this paper.

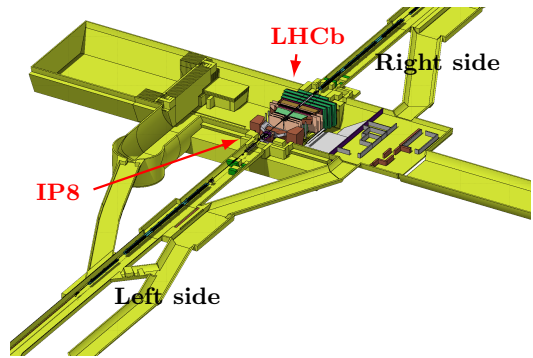
\begin{figure}[h!]
      \centering
       \begin{overpic}[trim=2.0cm 0.5cm 0.5cm 0.0cm, clip=true,width=0.8\linewidth]{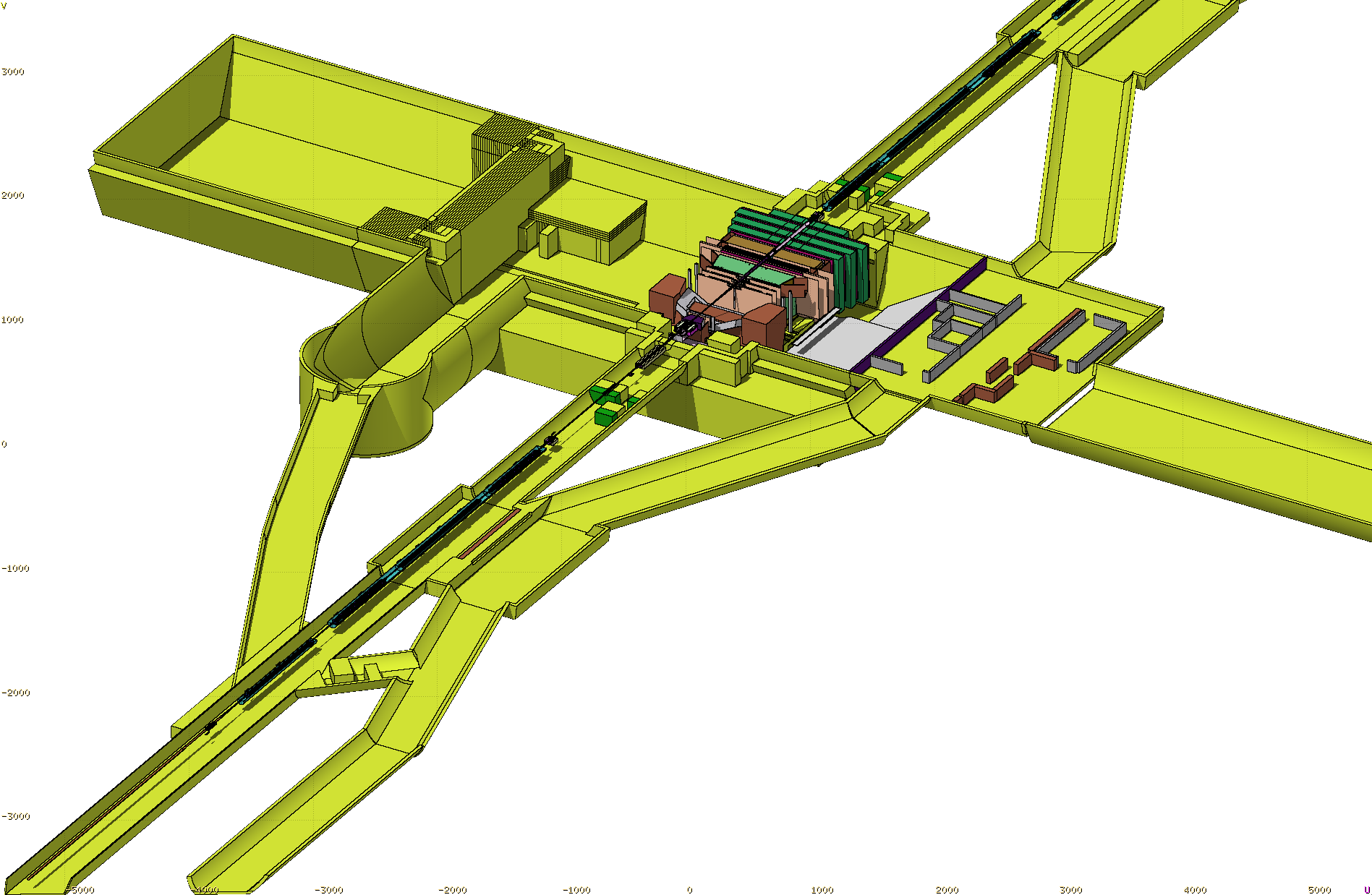}
        \put (500,600) {\textbf{\color{red} LHCb}}
        \put(550,580){\linethickness{0.3mm}\color{red}\vector(-0.1,-0.9){5}}
        \put (50,300) {\textbf{\color{red} IP8}}
        \put(160,320){\linethickness{0.3mm}\color{red}\vector(0.9,0.31){280}}
        \put (200,100) {\textbf{\color{black} Left side}}
        \put (700,550) {\textbf{\color{black} Right side}}
        \end{overpic}
      \caption{3D top view of the FLUKA geometry of IR8 including the LHCb detector, LHC tunnel and service areas.}
      \label{fig:3Dview_tot}
\end{figure} 
\begin{figure*}[!ht]
\setlength{\unitlength}{1\textwidth}
      \includegraphics[trim=0.0cm 0.3cm 0.0cm 0.0cm, clip=true,width=0.95\textwidth]{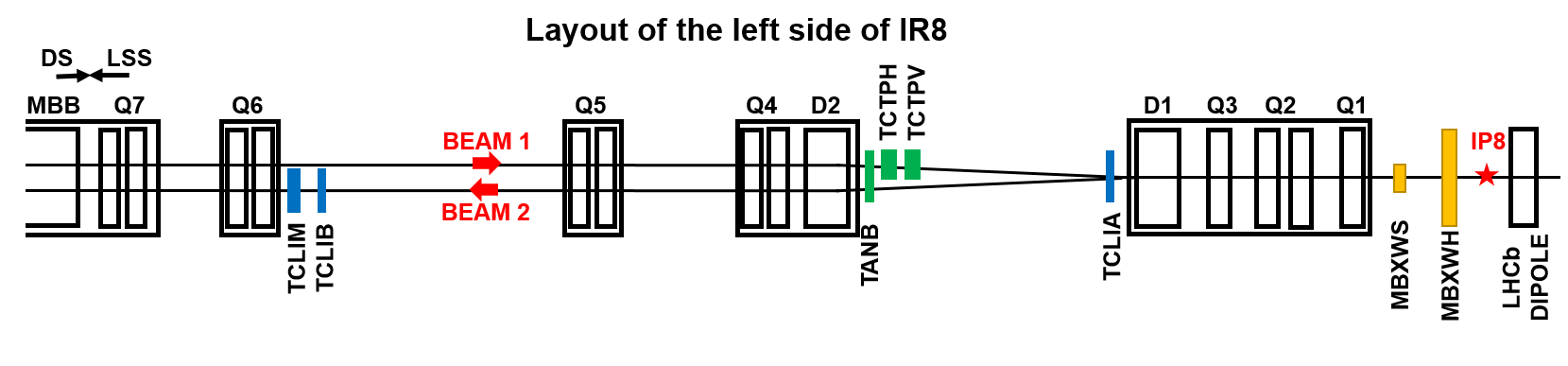}
      \includegraphics[trim=0.0cm 0.3cm 0.0cm 0.0cm, clip=true,width=0.95\textwidth]{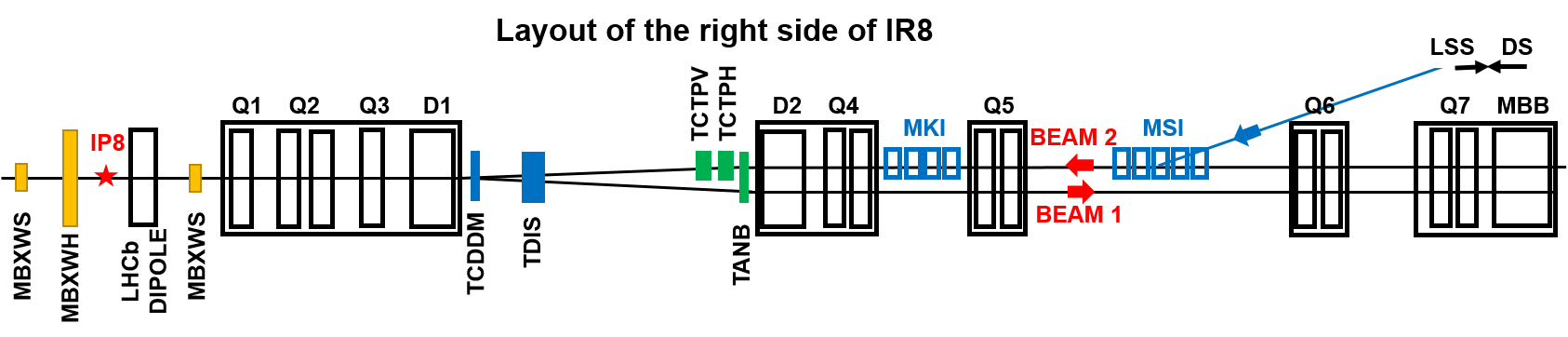}      
      \caption{Schematic layout of both sides of the IR8 Long Straight 
      Section (LSS) in Run~3. The blue elements (magnets and protection absorbers) are related to beam-2 injection. The yellow elements are the warm compensator magnets. The green elements are tertiary collimators and D2 protection absorbers (TANB), as listed in Table~\ref{tab:table1} and Table~\ref{tab:table2}}.
      \label{fig:layout}
\end{figure*} 
Our region of interest is presented in Section~\ref{sec:model}. In Section~\ref{sec:source}, a detailed characterization of the radiation source, namely the collision debris generated by inelastic collisions at IP8 and propagating along the final focus triplet and the separation dipole, is given. The validation of the model, based on BLM measurements collected during Run~2, is illustrated in Section~\ref{sec:run2}. Finally, in Section~\ref{sec:run3}, the findings for the upcoming Run~3 are reported, quantifying the exposure of the warm compensator magnets, cold quadrupoles, separation dipole, and recombination dipole, in order to assess quench risk and lifetime prospects.

\section{Simulation model: the LHC\lowercase{b} detector and the IR8 layout}
\label{sec:model}

The LHCb detector is designed to perform an indirect search of New Physics beyond the Standard Model. 
For this purpose, the experimental apparatus consists of a single-arm spectrometer optimized to work in the forward direction where the largest $b$ and $c$ quarks production is expected~\cite{Alves:1129809}. In order to conduct specific measurements, the detector doesn't surround the IP as in ATLAS and CMS, but it is spread over 20~metres of stacked sub-detectors placed only on the right-side of IP8 and just close to the beam. In order to exploit the space for the detector in the UX85 cavern where the DELPHI experiment had previously been housed, IP8 is displaced by approximately $11\,\mathrm{m}$ towards IP7, as shown in Fig.~\ref{fig:3Dview_tot}.
Given the difficulty to perform precise measurements in presence of too many primary vertexes, the detector was designed to operate at a lower luminosity than the two general purpose LHC experiments (ATLAS and CMS). The successive upgrades will allow much more data to be handled. 
\begin{table}[!hb]%The best place to locate the table environment is directly after its first reference in text
\caption{Protection elements on the left side of IP8.}
\label{tab:table1}
\begin{ruledtabular}
\begin{tabular}{ccc}
%\textrm{Left\footnote{Note a.}}&
\textrm{Element}&
\textrm{Distance from IP8 [m]}&
\textrm{Protection from}\\
\colrule
TCLIA.4L8 & -73 & \textrm{Injected beam B2}\\
TCTPV.4L8 & -116 & \textrm{Incoming beam B1}\\
TCTPH.4L8 & -118 & \textrm{Incoming beam B1}\\
TANB.4L8\footnote{Installed during LS2.} & -119 & \textrm{Collision debris}\\
TCLIB.6L8 & -217 & \textrm{Injected beam B2}\\
TCLIM.6L8 & -223 & \textrm{Injected beam B2}\\
\end{tabular}
\end{ruledtabular}
\end{table}

\begin{table}[!hb]%The best place to locate the table environment is directly after its first reference in text
\caption{Protection elements on the right side of IP8.}
\label{tab:table2}
\begin{ruledtabular}
\begin{tabular}{ccc}
%\textrm{Left\footnote{Note a.}}&
\textrm{Element}&
\textrm{Distance from IP8 [m]}&
\textrm{Protection from}\\
\colrule
TCDDM.4R8 & 71 & \textrm{Injected beam B2}\\
TDIS.4R8\footnote{Installed during LS2, replacing TDI.4R8.} & 81 & \textrm{Injected beam B2}\\
TCTPV.4R8 & 116 & \textrm{Incoming beam B2}\\
TCTPH.4R8 & 118 & \textrm{Incoming beam B2}\\
TANB.4R8\footnote{Installed during LS2.} & 119 & \textrm{Collision debris}\\
\end{tabular}
\end{ruledtabular}
\end{table}
The asymmetry with respect to the IP is a peculiarity of IR8 that makes it unique compared to the most studied IR1 and IR5. The shift of IP8 implies that the whole string of the quadrupole triplet and the separation and recombination dipoles is displaced by approximately $11\,\mathrm{m}$ towards the left side with respect to the center of the experimental cavern (that is also the magnetic center of the octant). The shift of IP8 is regained before the Dispersion Suppressor (DS), resulting in an asymmetric layout of the matching section with respect to IP8.

The LHCb spectrometer induces a horizontal kick of $194\,\mathrm{\mu}$rad on a $6.5~\mathrm{TeV}$ proton. Depending on its powering, the kick is directed towards the outside or the inside of the ring.
\begin{figure}[h!]
      \centering
       \begin{overpic}[trim=0.0cm 0.3cm 0.0cm 0.0cm, clip=true,width=1.0\linewidth]{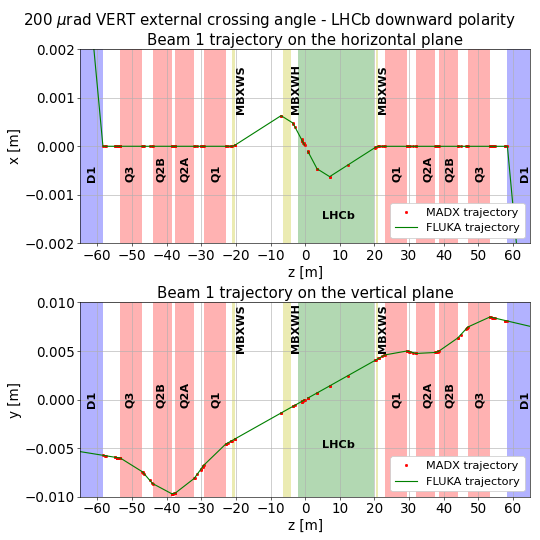}
        \end{overpic}
      \caption{Beam-1 trajectory between the left and right separation dipole (D1) of IR8, in the horizontal (top) and vertical (bottom) plane, for the reference $7\,\mathrm{TeV}$ proton. The green line shows the trajectory calculated with FLUKA and red points are extracted from the optics file generated with MAD-X.}
      \label{fig:traj_B1}
\end{figure} 
This orbit bump is compensated by a system of three warm dipoles, shown in yellow in the schematic layouts in Fig.~\ref{fig:layout}. The long warm dipole MBXWH is placed on the left side of IP8, as symmetric counterpart of the LHCb spectrometer giving an opposite kick, which turns out to be identical when considering two protons leaving IP8 in opposite directions. Two short compensators are placed just on the IP side of the (left and right) triplet, producing each a small kick of $49\,\mathrm{\mu}$rad opposite to the larger one of the MBXWH or spectrometer close by. On the whole, a $6.5~\mathrm{TeV}$ proton traveling from one triplet to the other through IP8 experiences a $\pm 49,\mp 194, \pm 194,\mp 49\,\mathrm{\mu}$rad kick sequence and so arrives in IP8 with an angle of $\mp145\,\mathrm{\mu}$rad on the horizontal plane. Since the field intensities of the dipoles are kept constant, this angle becomes $\mp 135\,\mathrm{\mu}$rad for $\sqrt{s}=14\,\mathrm{TeV}$, as shown in the top trajectory plot of Fig.~\ref{fig:traj_B1}.

The actual angle between the proton momentum in IP8 and the detector axis is determined by the combination of the above bump with the external crossing angle. The latter is required to prevent undesired encounters in the region where the beams share the same vacuum chamber, and it is enabled by corrector dipoles~\cite{Herr:1159131}. Several configurations with different crossing plane and crossing angle value have been studied and adopted during operation. As an example, at the end of Run~2 in 2018 the external crossing angle was on the horizontal plane. However, for symmetry reasons, the choice of an external crossing angle on the vertical plane is preferred by LHCb, which accumulates equal integrated luminosities with either spectrometer polarity~\cite{Albrecht:2653011}. Therefore, during Run~3, from 2023 onwards, the external crossing angle is planned to be enabled on the vertical plane, resulting in a skew crossing plane in IP8~\cite{Fartoukh:2790409} (see the bottom plot of Fig.\ref{fig:traj_B1}).

The injection line of the counterclockwise beam (beam~2) joins the LHC on the right side of IR8, implying the presence of septa magnets in the half-cell~6 and kicker magnets in the half cell~5, which are switched off during physics production (see Fig.~\ref{fig:layout}). In addition, dedicated injection protection elements are in place, namely the TCLIA and TCLIB collimators on the left side (see Table~\ref{tab:table1}) and the TDI absorber and TCDDM mask on the right side (see Table~\ref{tab:table2}). During LS2, the LHC injection protection system was upgraded by the replacement of the TDI with a new segmented absorber, called TDIS~\cite{Uythoven:IPAC2015-TUPTY051}. 

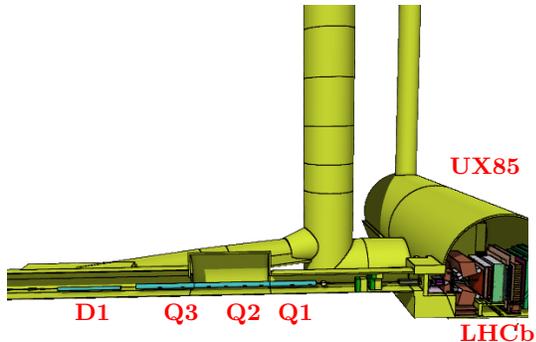
\begin{figure}[h!]
      \centering
       \begin{overpic}[trim=2.0cm 0.5cm 0.5cm 0.0cm, clip=true,width=0.8\linewidth]{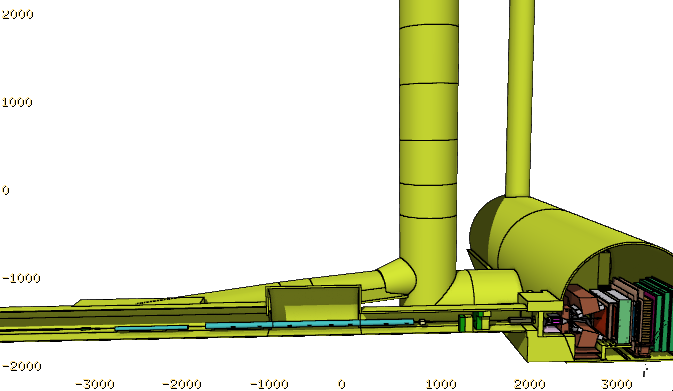}
        \put (870,-20) {\textbf{\color{red} LHCb}}
        \put (850,300) {\textbf{\color{red} UX85}}
        \put (520,20) {\textbf{\color{red} Q1}}
        \put (420,20) {\textbf{\color{red} Q2}}
        \put (300,20) {\textbf{\color{red} Q3}}      
        \put (130,20) {\textbf{\color{red} D1}}   
        \end{overpic}
      \caption{3D view (from the inside of the ring) of the FLUKA geometry including the LHCb experiment and the left side of the LHC with the quadrupole triplet Q1-Q3 and the separation dipole D1.}
      \label{fig:3Dview}
    \end{figure}

Regarding additional machine protection elements, two tertiary collimators per side (TCTPH and TCTPV) are installed to protect the experiment from the incoming beams (see Fig.~\ref{fig:layout}, Table~\ref{tab:table1} and Table~\ref{tab:table2}). Otherwise, recalling the fact that LHCb was designed to operate at a lower luminosity than ATLAS and CMS, the TAS (Target Absorber Secondaries) and TAN (Target Absorber Neutrals) absorbers and TCL (Target Collimator Long) collimators for physics debris were not necessary in IR8 for Run~1 and Run~2. The TAS is an absorber placed on the IP side of the final focus triplet to protect the first quadrupole (Q1) from the collision debris, while the TAN is an absorber intercepting mostly neutral particles, being placed in front of the recombination dipole (D2) where there is the transition between the common vacuum chamber and the two separate beam pipes. Nevertheless, during LS2 a new TANB (Target Absorber Neutrals at LHCb region) absorber was installed at 120 m from IP8 on both sides, in view of the luminosity increase foreseen for Run~3~\cite{1960537,1961576}. On the contrary, the TAS absorber and TCL collimators are still not required in IR8 for the planned operation up to $2\cdot10^{33}\,\mathrm{cm}^{−2}\,\mathrm{s}^{−1}$, aiming to produce up to $50\,\mathrm{fb}^{−1}$.
Two FLUKA models of the detector, corresponding to the Run~2 and Run~3 version, respectively, have been included in the FLUKA repository as developed by the LHCb Collaboration~\cite{Karacson:2243499}. The Run~3 geometry includes major upgrades in key detector elements, like the neutron shielding.
Figure~\ref{fig:3Dview} gives another 3D view of the comprehensive geometry, with the shafts to the surface. 
In order to implement this geometry, a Python-based tool for assembling accelerator beam lines (e.g., LHC, SPS, PS) for FLUKA simulations, called Linebuilder, has been used \cite{Mereghetti:2012zz}. The software interfaces with a library, the FLUKA Element Database, including the FLUKA geometry models of different accelerator components (magnets, collimators, absorbers, etc.), which are used with a modular approach to build the beam line on the basis of optics (Twiss) files.

At the same time, the BLMs~\cite{HOLZER20122055,PhysRevAccelBeams.22.071003} are placed according to the positions extracted from the CERN Layout database~\cite{CERN_Layout_Database} as well as corrected by visual inspection in the tunnel. For the benchmarking studies, lower transport thresholds have been applied as described in~\cite{PhysRevAccelBeams.22.071003}.
The beam trajectory over the considered region, as simulated in FLUKA, is consistent with the nominal optics within an accuracy of few\,$\mu$m (see Fig.~\ref{fig:traj_B1}).

\section{Source term: collision debris}
\label{sec:source}
For proton operation, radiation showers in the experimental IRs are dominated by inelastic nuclear interactions at the IP. Hence, this study doesn't include elastic interactions, whose products remain in the beam and are possibly intercepted by the collimation system~\cite{Brugger:2131739}. Each inelastic collision at the IP may generate a large number of secondary particles, on average about 120 with $7~\mathrm{TeV}$ beams. Due to decay of unstable particles, mainly neutral pions, already 5 mm away from the IP and without interacting with any material, the number of debris particles increases to about 155, of which $50\%$ are photons and $35\%$ are charged pions~\cite{Brugger:2131739}. While the majority of these particles interacts in the experimental beam pipe and in the detector , the most energetic debris is scattered at small angles with respect to the beam direction. These particles propagate along the beam line in the IRs, impacting on the machine elements and determining radiation levels in the LHC tunnel and in the nearby locations. 
\begin{figure}[!h]
      \centering
         \centering
         \begin{overpic}[trim=0.0cm 0.0cm 0.0cm 0.0cm, clip=true,width=1\linewidth]{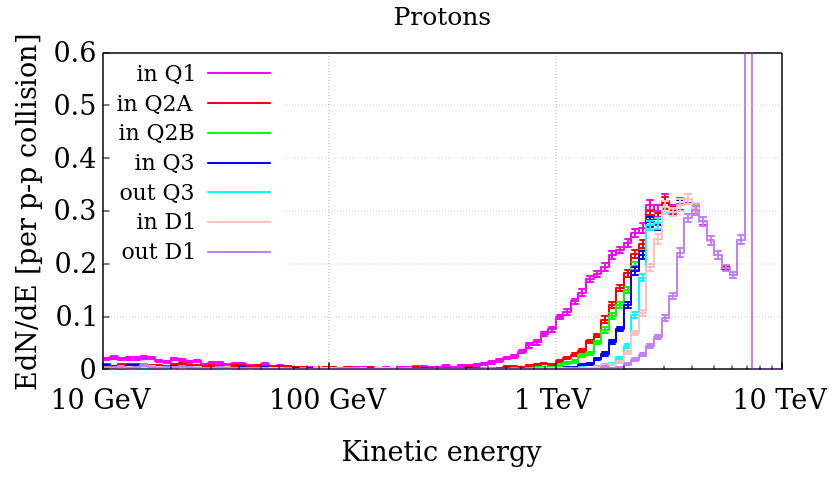}
         \end{overpic}
         \label{fig:proton}
      \centering
      \begin{overpic}[trim=15.0cm 0.0cm 0.0cm 0.0cm, clip=true,width=0.8\linewidth]{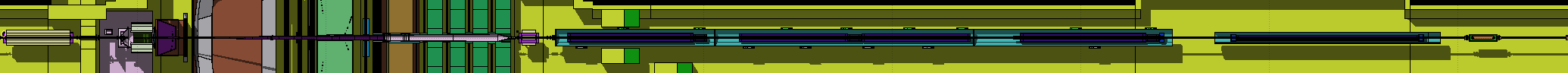}
      \put(120,0){\linethickness{0.3mm}\color{magenta}\vector(0,1){5}}
      \put(280,0){\linethickness{0.3mm}\color{red}\vector(0,1){5}}
      \put(400,0){\linethickness{0.3mm}\color{green}\vector(0,1){5}}
      \put(520,0){\linethickness{0.3mm}\color{blue}\vector(0,1){5}}
      \put(650,0){\linethickness{0.3mm}\color{cyan}\vector(0,1){5}}
      \put(700,0){\linethickness{0.3mm}\color{pink}\vector(0,1){5}}
      \put(900,0){\linethickness{0.3mm}\color{violet}\vector(0,1){5}}
      \end{overpic}
    \caption{Energy spectra in lethargy unit, of protons in the vacuum pipe of the right side of IP8 at the position indicated by the arrows in the geometry layout.}
    \label{fig:p}
\end{figure}
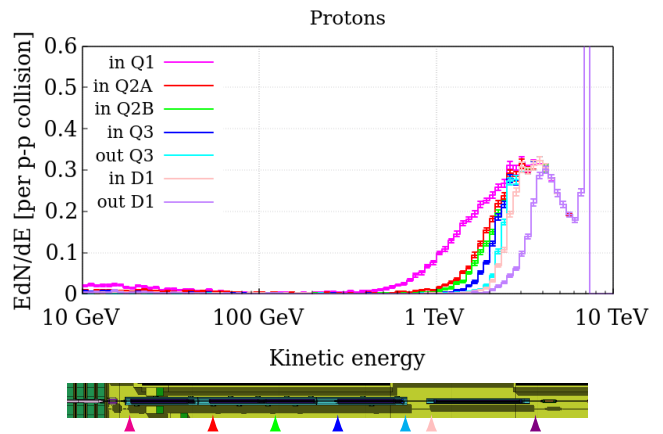 
The power absorbed by the final focus quadrupoles and the separation  dipole is mostly due to charged pions that are bent by the magnetic field onto the beam screen walls representing the mechanical aperture.
    \begin{figure}[h!]
      \centering
         \centering
         \includegraphics[width=1\linewidth]{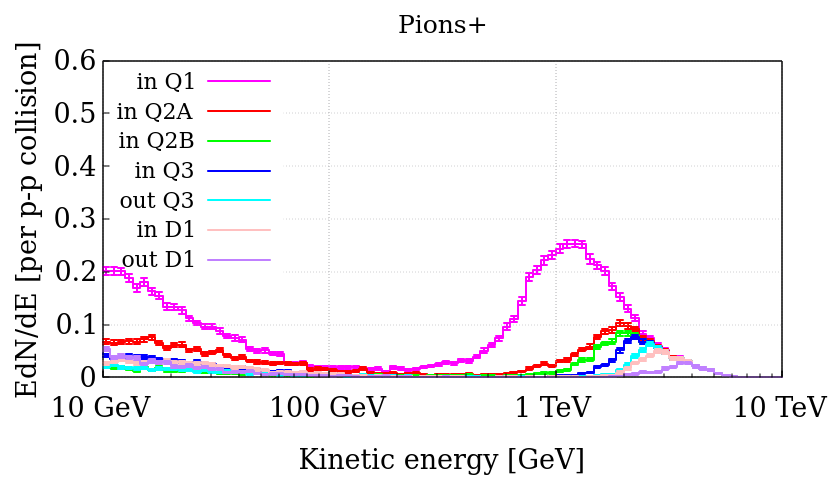}
      \label{fig:pion+}
         \centering
         \includegraphics[width=1\linewidth]{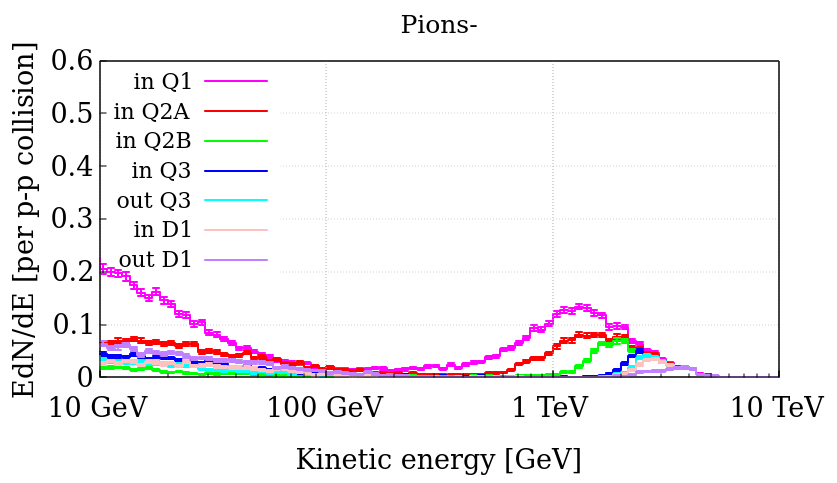}
      \label{fig:pion-}
      \centering
      \begin{overpic}[trim=15.0cm 0.0cm 0.0cm 0.0cm, clip=true,width=0.8\linewidth]{plot/geo_1.png}
      \put(120,0){\linethickness{0.3mm}\color{magenta}\vector(0,1){5}}
      \put(280,0){\linethickness{0.3mm}\color{red}\vector(0,1){5}}
      \put(400,0){\linethickness{0.3mm}\color{green}\vector(0,1){5}}
      \put(520,0){\linethickness{0.3mm}\color{blue}\vector(0,1){5}}
      \put(650,0){\linethickness{0.3mm}\color{cyan}\vector(0,1){5}}
      \put(700,0){\linethickness{0.3mm}\color{pink}\vector(0,1){5}}
      \put(900,0){\linethickness{0.3mm}\color{violet}\vector(0,1){5}}
        \end{overpic}
      \caption{Energy spectra in lethargy unit, of positive pions (on the top plot) and of negative pions (on the bottom plot) in the vacuum pipe of the right side of IP8 at the position indicated by the arrows  in the geometry layout.}
      \label{fig:pions}
    \end{figure}
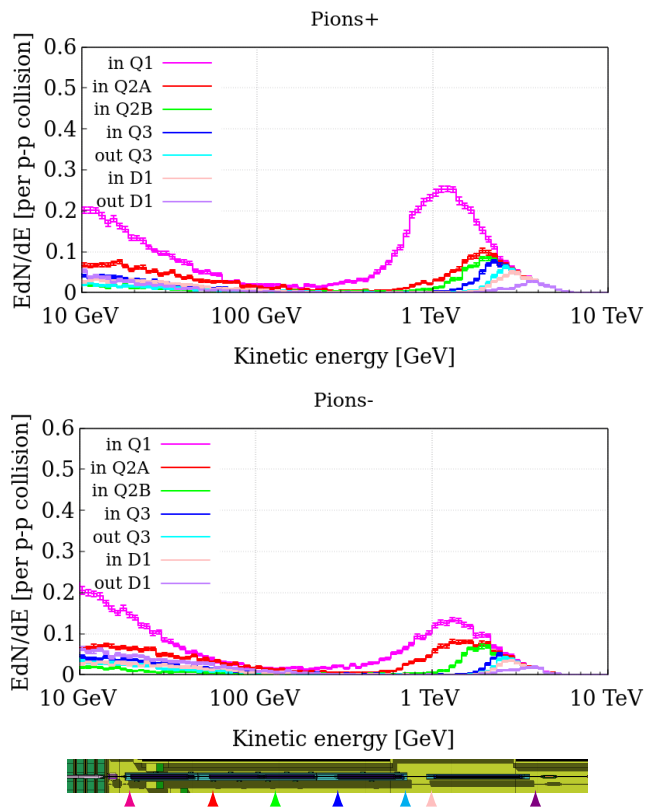 
For a luminosity of $4\cdot10^{32}\,\mathrm{cm}^{−2}\,\mathrm{s}^{−1}$, this corresponds to $32\cdot10^6$ inelastic collisions per second and a power of 33~W towards either (right or left) side.

In this work, the inelastic cross section, including diffractive events, is assumed to be $\sigma_{pp}=80\,\mathrm{mb}$. In Figs~\ref{fig:p}-\ref{fig:pions} the energy spectra of charged particles travelling inside the vacuum chamber(i.e., protons and charged pions that are the dominant species) are shown at different positions along the triplet and the separation dipole.

Their behaviour along the accelerator line is influenced both by the initial conditions of the collision, i.e. the beam crossing scheme, and the magnetic fields which they are subject to. As for the triplet, the configuration of the four quadrupoles (Q1, Q2A, Q2B, Q3) is DFFD (defocusing-focusing-focusing-defocusing) in the horizontal plane for the outgoing beam. The particles with a lower magnetic rigidity than the circulating beam may be captured. As an example, considering the case of $\sqrt{s}=14\,\mathrm{TeV}$, only protons with an energy higher than $5.5\,\mathrm{TeV}$ can reach up to $200\,\mathrm{m}$ from IP8, while a $5\,\mathrm{TeV}$ proton is lost before the recombination dipole.

The $7\,\mathrm{TeV}$ peak in Fig.~\ref{fig:p} is due to diffractive processes, where one interacting proton receives a limited angular kick and is subject to a slight energy loss, this way managing to leave the Long Straight Section (LSS).
\begin{figure}[h!]
      \centering
      \includegraphics[trim=3.0cm 0.0cm 4.0cm 0.0cm, clip=true,width=1\linewidth]{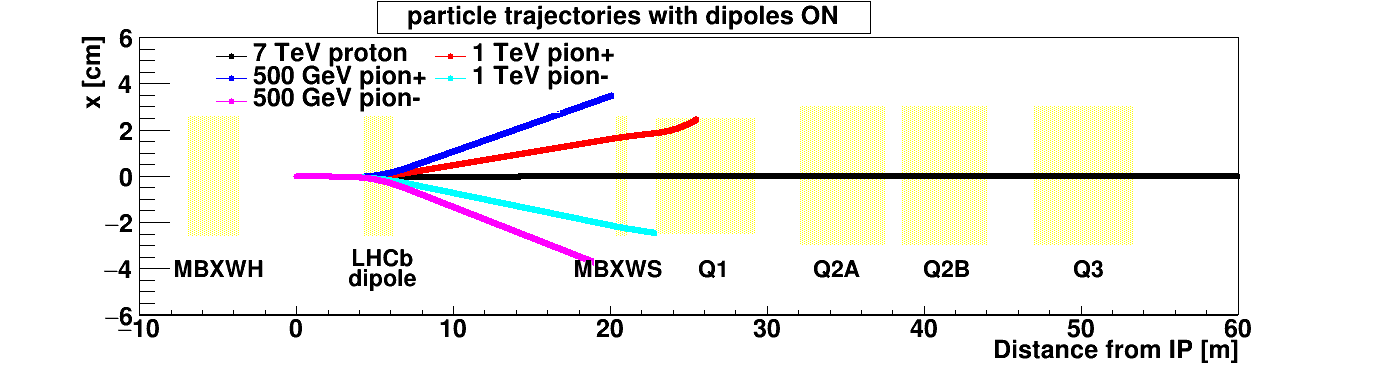}
      \includegraphics[trim=3.0cm 0.0cm 4.0cm 0.0cm, clip=true,width=1\linewidth]{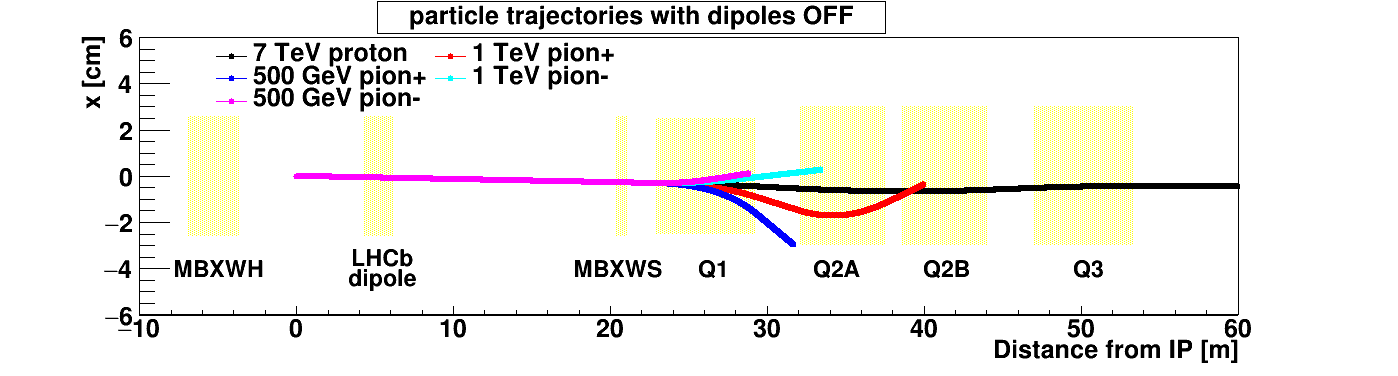}
      \caption{Beam proton and secondary pion trajectories in the horizontal plane on the right side of IP8, for external crossing in the vertical plane and downward LHCb polarity. Yellow areas indicate the magnet mechanical aperture. In the bottom panel, the LHCb spectrometer and short compensator fields are academically zeroed, in order to appreciate their effect. The pink, cyan and red trajectories end where they meet the beam screen wall in the vertical plane.}
      \label{fig:compensator_effect}
\end{figure} 
In the range between $500\,\mathrm{GeV}$ and $2\,\mathrm{TeV}$, the fluence at the IP-face of the Q1 in IR8 is higher than the values published for IR1 and IR5~\cite{Brugger:2131739} due to the absence of the TAS absorber. As already introduced, the TAS absorber, installed in the high luminosity IRs, protects the first quadrupole and considerably reduces the absorbed power as well as the peak dose and power density in its coils.
The pion spectra of Fig.~\ref{fig:pions} are peaked at lower energies just above $1\,\mathrm{TeV}$ on the Q1 front face, and their average energy increases for larger distances from IP8 as the smaller magnetic rigidity component gets captured along the way. Most pions, especially positively charged, are captured by the magnetic field of the Q1 quadrupole. 

The pink tail below $100\,\mathrm{GeV}$ comes from debris re-interactions in the experimental cavern. 
As for the high energy part, positive pions are significantly more abundant than negative pions, because a larger fraction of the latter ones does not even reach the triplet due to the combined effect of the crossing angle and the LHCb spectrometer (or MBXWH) field~\cite{LHCbUpgrade2020_Cerutti,Efthymiopoulos:2319258}. In fact, the top panel of Fig.~\ref{fig:compensator_effect} shows that negative particles are further bent on the same side pointed to by the crossing angle (at negative $x$ for the considered LHCb polarity) and so miss the Q1 aperture. The role of the LHCb and MBXWH field is emphasized by the comparison with the bottom panel, where no field is applied before entering the triplet.
We note that this difference in the abundance of positive and negative pions is much less dramatic in IR1 and IR5~\cite{Brugger:2131739}. The low energy tail of the pink curve in Fig.~\ref{fig:pions}, is not present in the case of IR1 and IR5 due to the shielding by the TAS. In addition, re-interactions in the TAS absorber itself cause the peak to be at a lower energy, namely around $500\,\mathrm{GeV}$, and more pronounced than that observed in IR8~\cite{Brugger:2131739}. 
	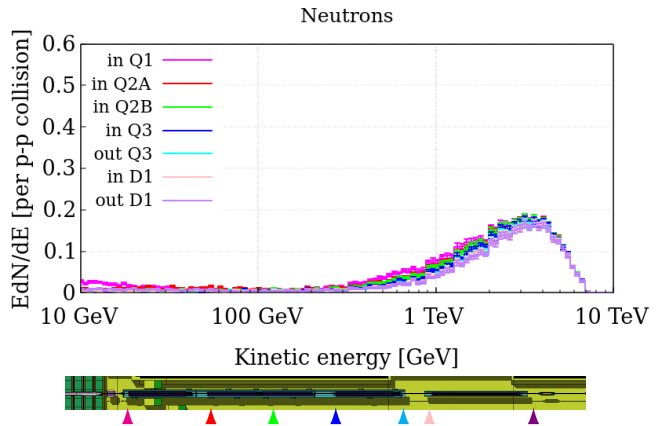
\begin{figure}[!h]
      \centering
 %       \begin{subfigure}[b]{0.8\linewidth}
         \centering
       \begin{overpic}[trim=0.0cm 0.0cm 0.0cm 0.0cm, clip=true,width=1\linewidth]{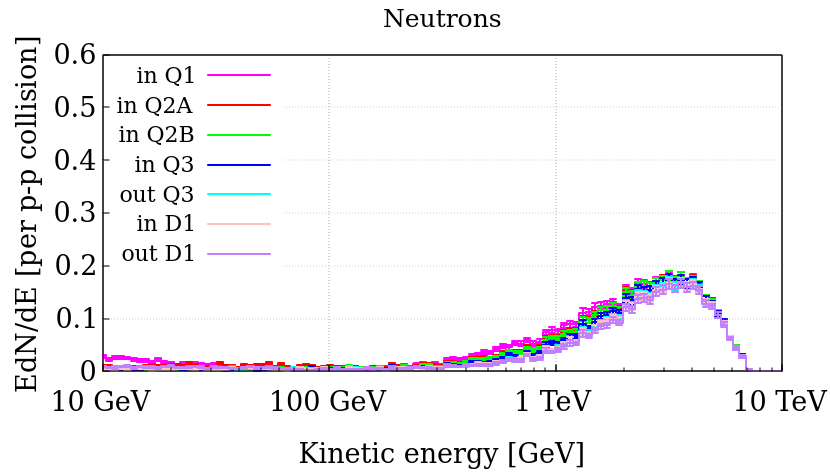}
%        \put (130,20) {\textbf{\color{red} D1}}   
        \end{overpic}
%      \includegraphics[width=0.8\textwidth]{plot/LHCbrun3_wall_rev1_7_exp_MoEDAL.png}
 %    \end{subfigure}

 %     \begin{subfigure}[b]{0.8\linewidth}
      \centering
      \begin{overpic}[trim=15.0cm 0.0cm 0.0cm 0.0cm, clip=true,width=0.8\linewidth]{plot/geo_1.png}
      \put(120,0){\linethickness{0.3mm}\color{magenta}\vector(0,1){5}}
      \put(280,0){\linethickness{0.3mm}\color{red}\vector(0,1){5}}
      \put(400,0){\linethickness{0.3mm}\color{green}\vector(0,1){5}}
      \put(520,0){\linethickness{0.3mm}\color{blue}\vector(0,1){5}}
      \put(650,0){\linethickness{0.3mm}\color{cyan}\vector(0,1){5}}
      \put(700,0){\linethickness{0.3mm}\color{pink}\vector(0,1){5}}
      \put(900,0){\linethickness{0.3mm}\color{violet}\vector(0,1){5}}
     %        \put (130,20) {\textbf{\color{red} D1}}   
        \end{overpic}
        
%      \end{subfigure}
      \caption{Energy spectra in lethargy unit, of neutrons in the vacuum pipe of the right side of IP8 at the position indicated by the arrows in the geometry layout.}
      \label{fig:neutron}
    \end{figure} 

	\begin{figure}[!h]
      \centering
         \centering
       \begin{overpic}[trim=0.0cm 0.0cm 0.0cm 0.0cm, clip=true,width=1\linewidth]{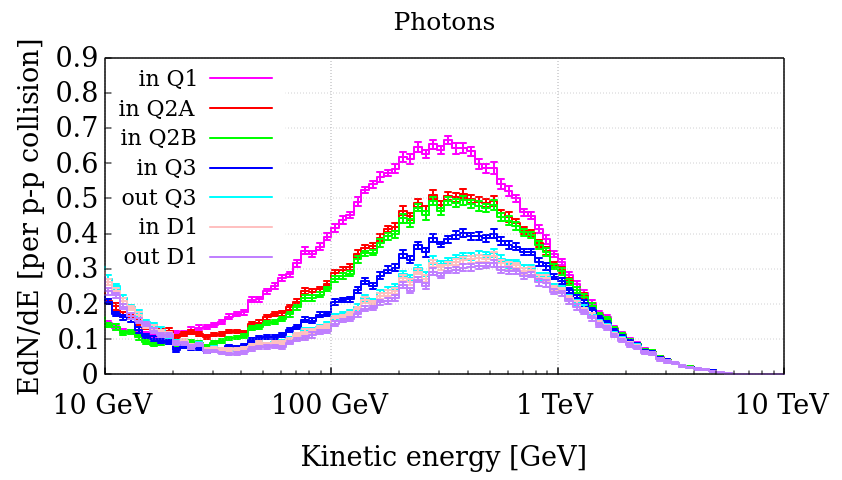}
        \end{overpic}
      \centering
      \begin{overpic}[trim=15.0cm 0.0cm 0.0cm 0.0cm, clip=true,width=0.8\linewidth]{plot/geo_1.png}
      \put(120,0){\linethickness{0.3mm}\color{magenta}\vector(0,1){5}}
      \put(280,0){\linethickness{0.3mm}\color{red}\vector(0,1){5}}
      \put(400,0){\linethickness{0.3mm}\color{green}\vector(0,1){5}}
      \put(520,0){\linethickness{0.3mm}\color{blue}\vector(0,1){5}}
      \put(650,0){\linethickness{0.3mm}\color{cyan}\vector(0,1){5}}
      \put(700,0){\linethickness{0.3mm}\color{pink}\vector(0,1){5}}
      \put(900,0){\linethickness{0.3mm}\color{violet}\vector(0,1){5}}
        \end{overpic}
      \caption{Energy spectra in lethargy unit, of photons in the vacuum pipe of the right side of IP8 at the position indicated by the arrows in the geometry layout.}
      \label{fig:photon}
    \end{figure} 

    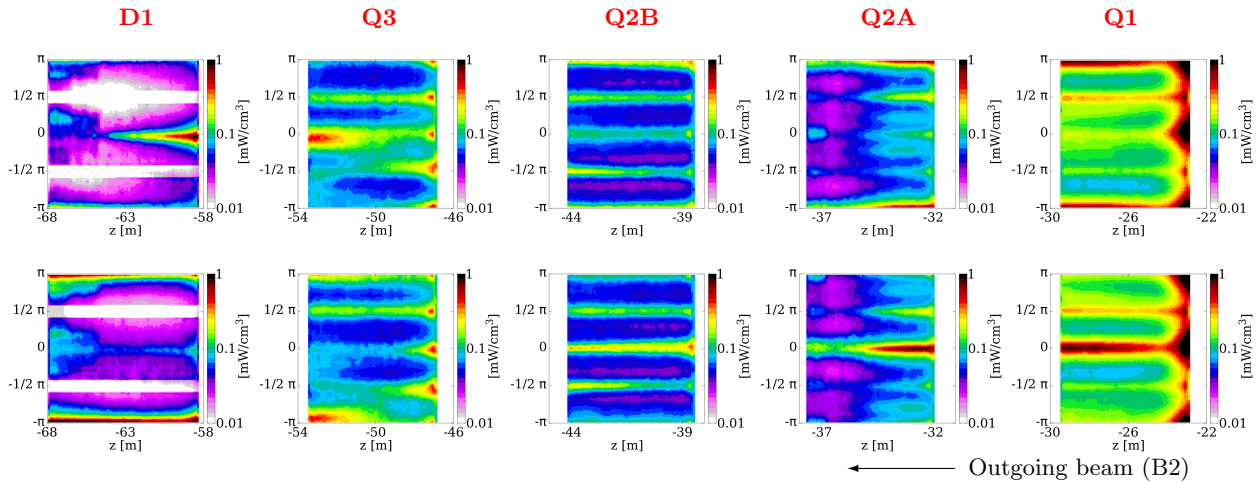
\begin{figure*}[!ht]
\setlength{\unitlength}{1\textwidth}
       \begin{picture}(1,0.03)

       \put (0.85,0.0005) {\textbf{\color{red} Q1}}
       \put (0.67,0.0005) {\textbf{\color{red} Q2A}}
        \put (0.48,0.0005) {\textbf{\color{red} Q2B}}
        \put (0.3,0.0005) {\textbf{\color{red} Q3}}      
        \put (0.12,0.0005) {\textbf{\color{red} D1}} 
       \end{picture}
       \includegraphics[trim=0.0cm 0.0cm 0.0cm 0.0cm,width=0.18\textwidth]{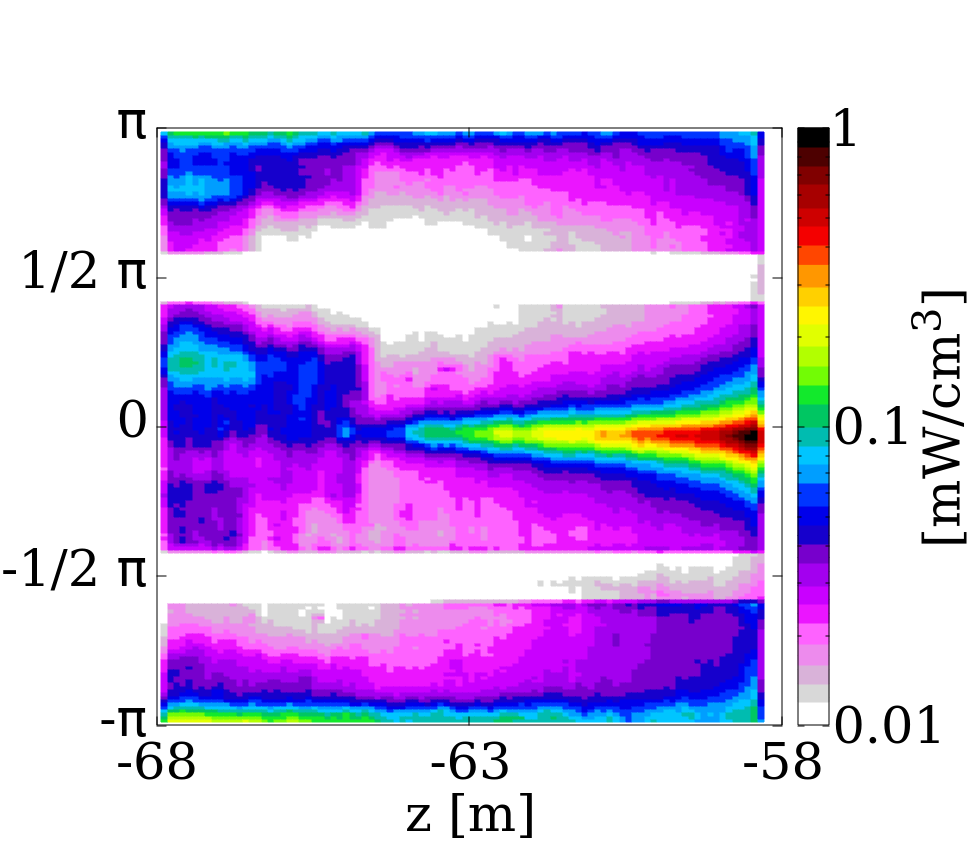}
       \includegraphics[trim=0.0cm 0.0cm 0.0cm
       0.0cm,width=0.18\textwidth]{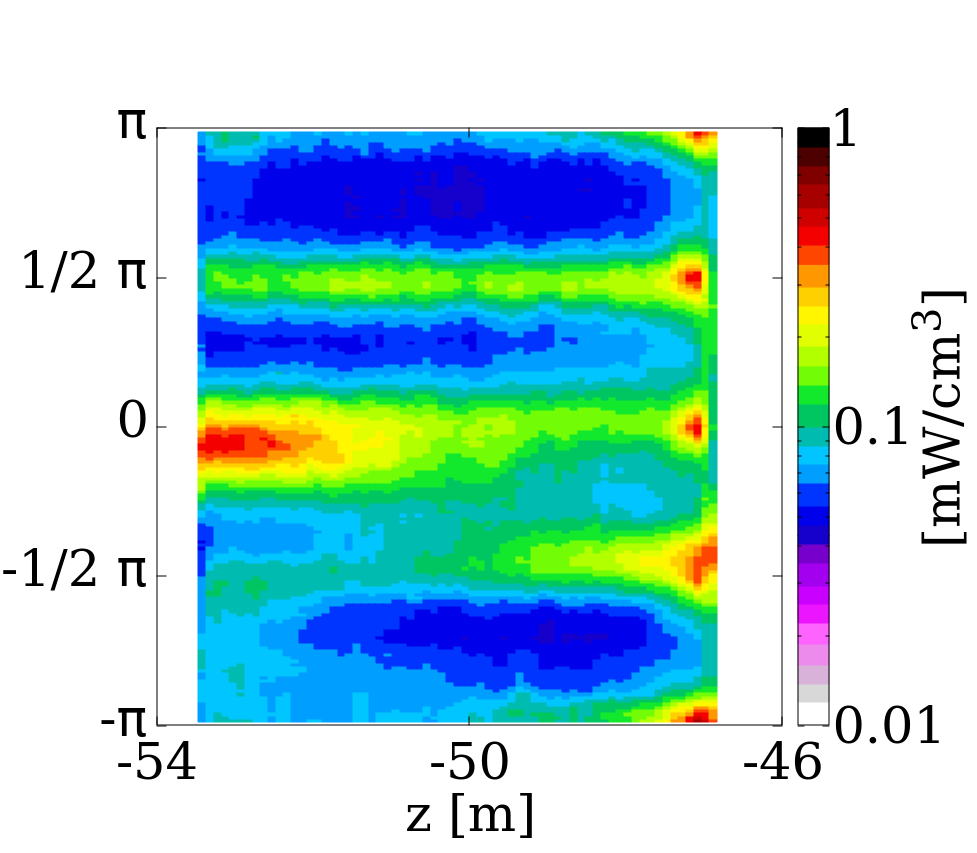}
       \includegraphics[trim=0.0cm 0.0cm 0.0cm 0.0cm,width=0.18\textwidth]{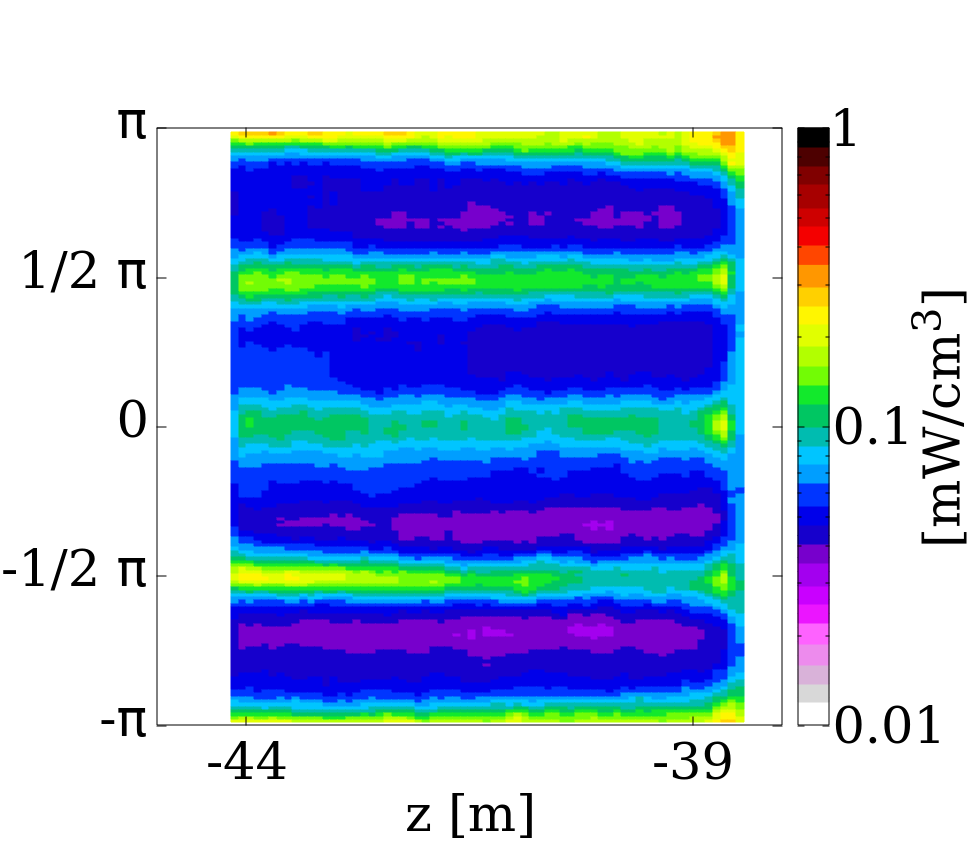}
       \includegraphics[trim=0.0cm 0.0cm 0.0cm 0.0cm,width=0.18\textwidth]{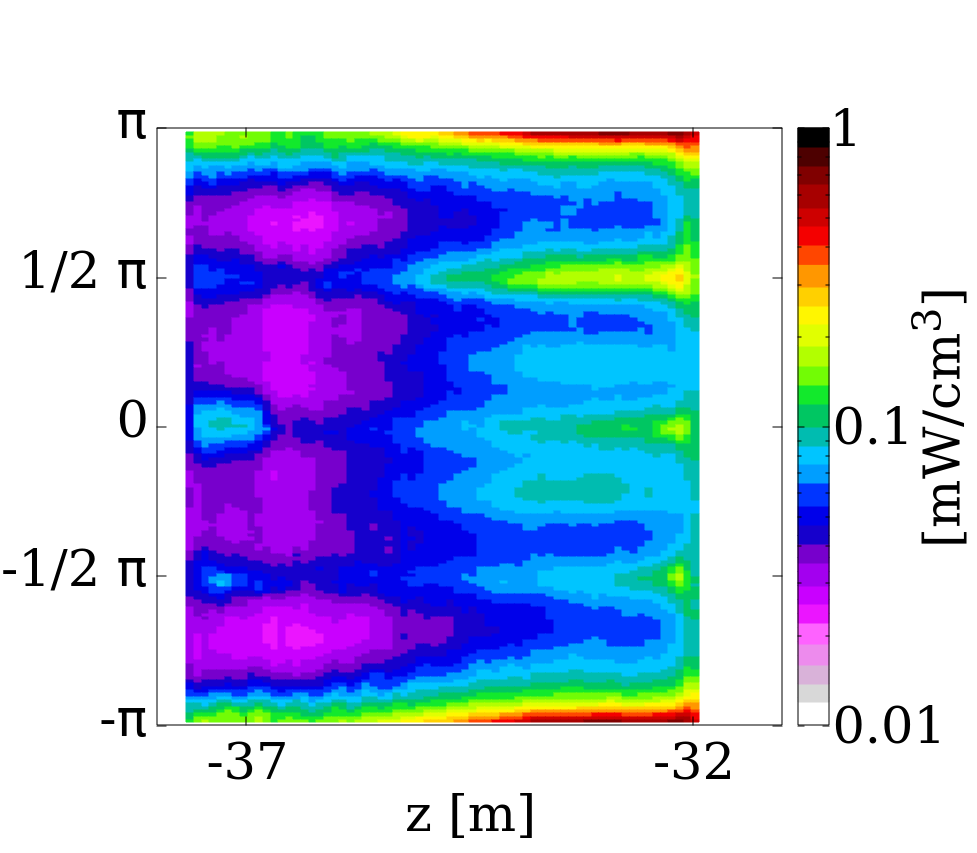}
       \includegraphics[trim=0.0cm 0.0cm 0.0cm 0.0cm,width=0.18\textwidth]{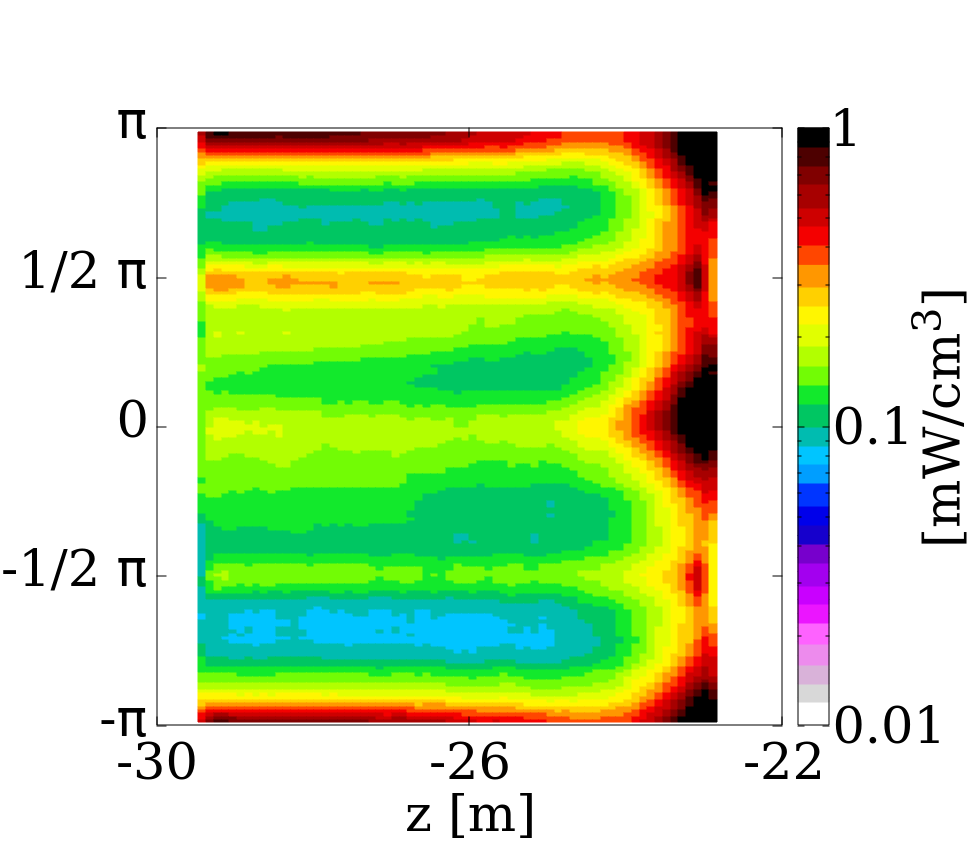}
       \includegraphics[trim=0.0cm 0.0cm 0.0cm 0.0cm,width=0.18\textwidth]{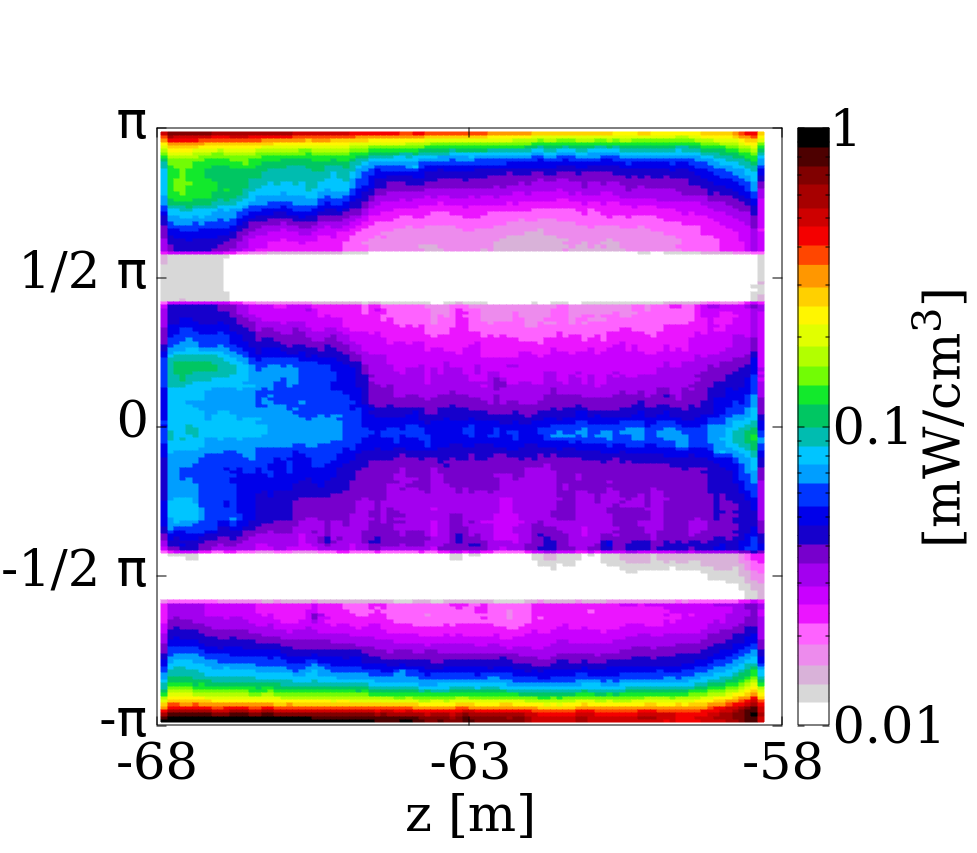}
       \includegraphics[trim=0.0cm 0.0cm 0.0cm
       0.0cm,width=0.18\textwidth]{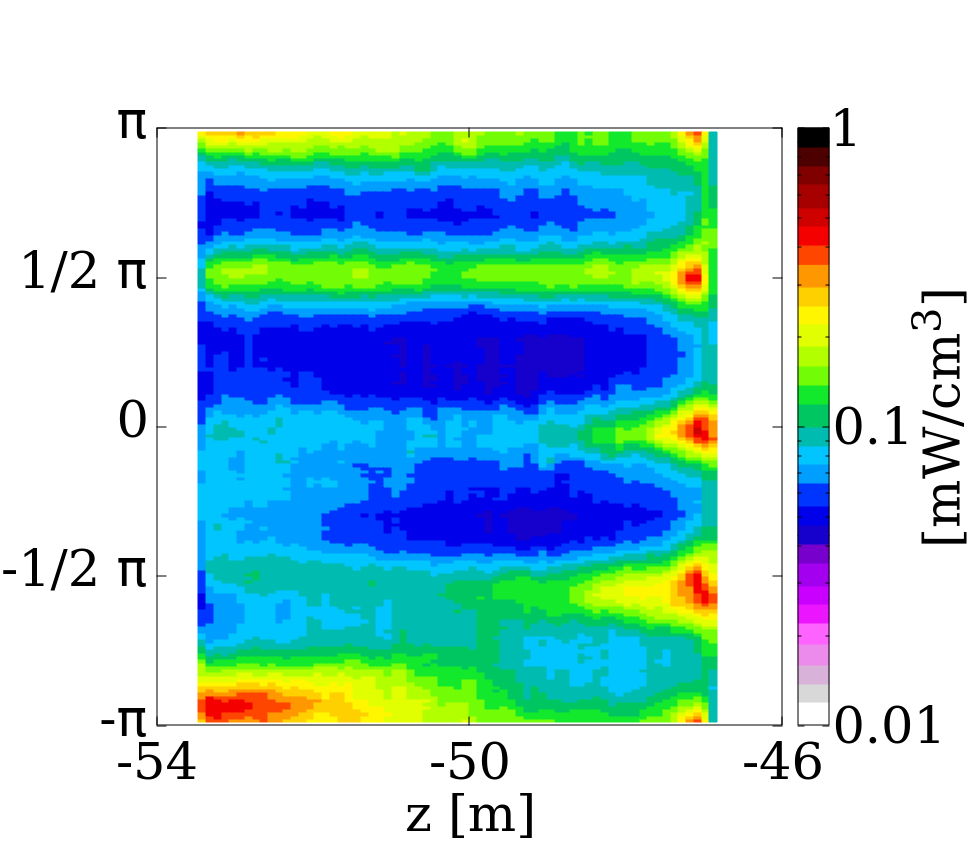}
       \includegraphics[trim=0.0cm 0.0cm 0.0cm 0.0cm,width=0.18\textwidth]{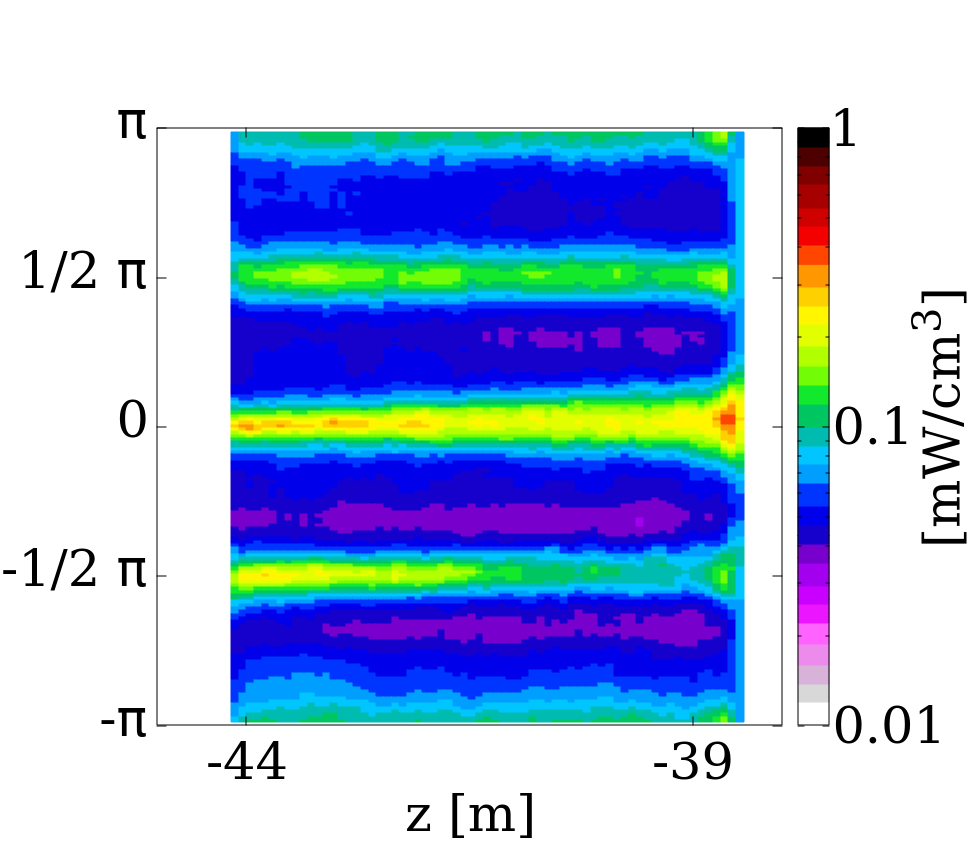}
       \includegraphics[trim=0.0cm 0.0cm 0.0cm 0.0cm,width=0.18\textwidth]{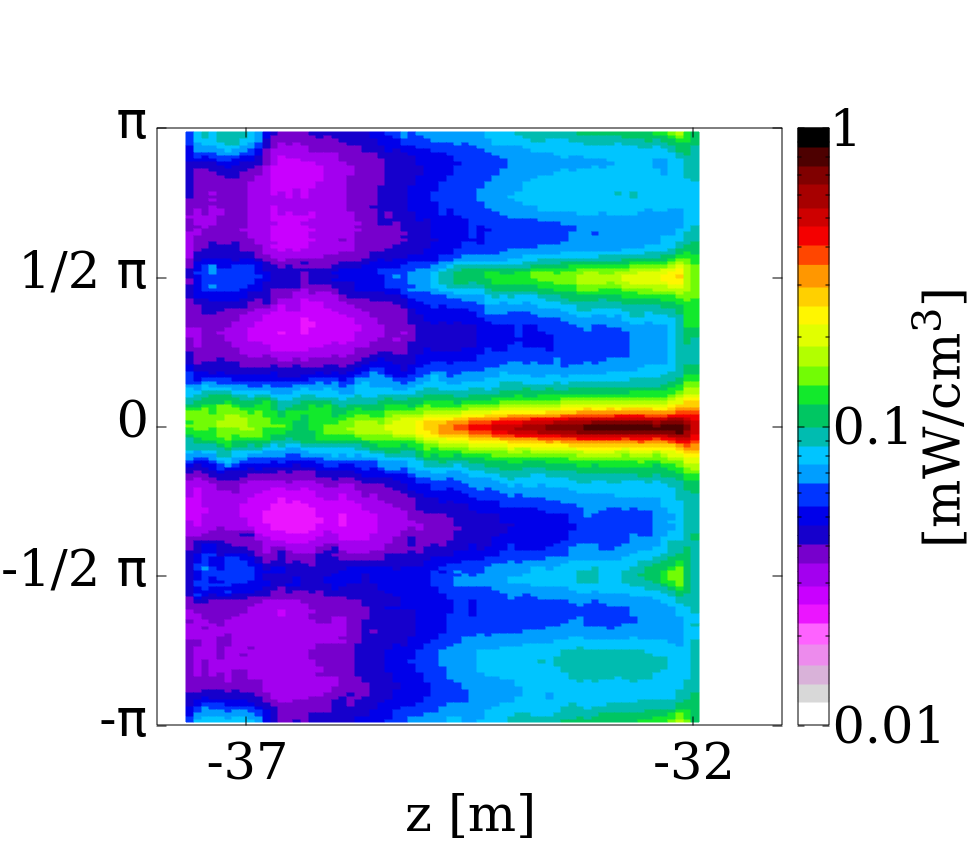}
       \includegraphics[trim=0.0cm 0.0cm 0.0cm 0.0cm,width=0.18\textwidth]{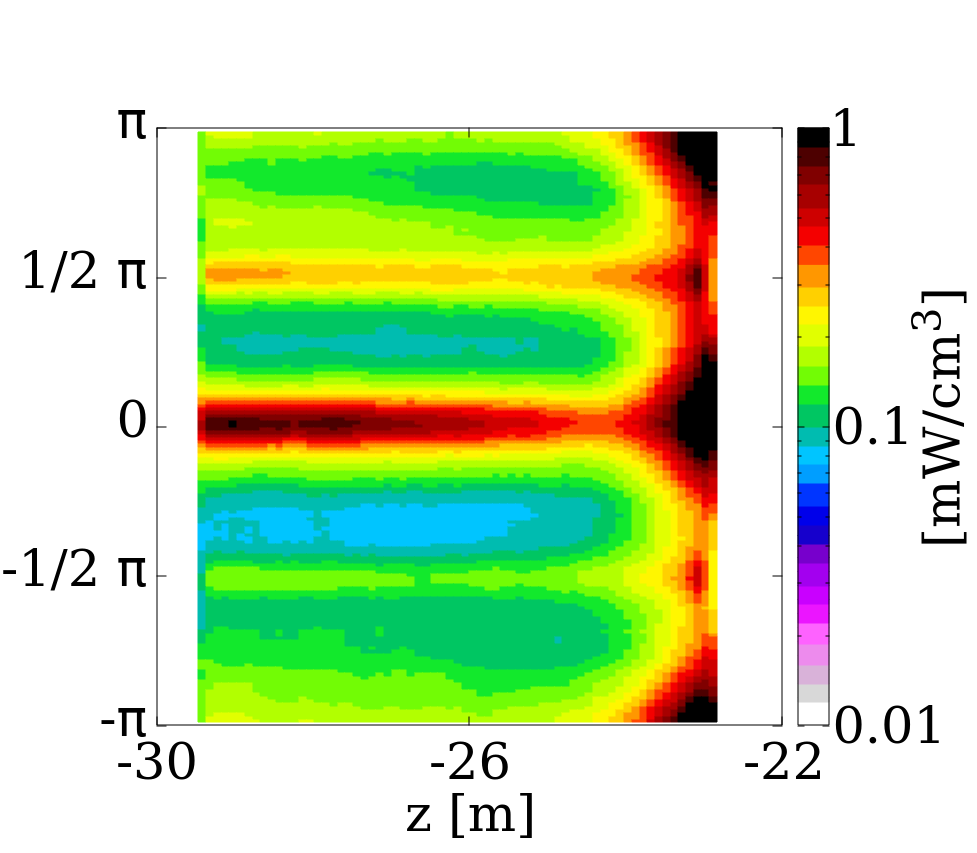}
       \begin{picture}(1,0.01)
       \put(0.74,0.01){\vector(-1,0){0.08}}
       \put(0.75,0.005){Outgoing beam (B2)}
       \end{picture}
       \caption{Colour maps of deposited power density as a function of the distance from IP8 (abscissa) and azimuthal angle (ordinate) for the upward (top plots) and downward (bottom plots) polarity of the LHCb spectrometer, with an external half crossing angle of $200\,\mathrm{\mu}$rad in the vertical plane. The plots refer, from right-to-left, to the four quadrupoles Q1-Q2A-Q2B-Q3 and the D1 separation dipole located on the left side of IP8. Power density values are averaged over the radial thickness of their inner coil layer and are given in $\mathrm{mW/cm^3}$, normalized to an instantaneous luminosity of $2\cdot10^{33}\,\mathrm{cm}^{−2}\,\mathrm{s}^{−1}$, for $7~\mathrm{TeV}$ proton beam operation. The azimuthal angle $\phi$ runs over the interval $(−\pi, +\pi]$ in radians, where $\phi=0$ is the horizontal direction pointing outside the ring, $\phi=\pi/2$ is the vertical direction opposite to gravity and $\phi=\pm\pi$ is the horizontal direction pointing inside the ring.}
       \label{fig:vert}
             
%       \end{subfigure}

\end{figure*} 
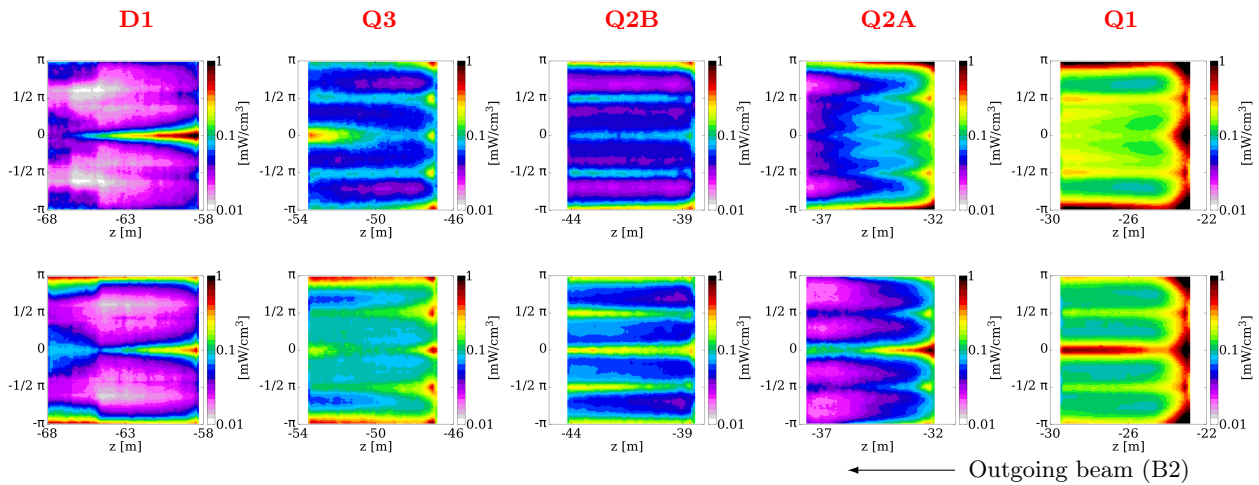
\begin{figure*}[!ht]
%       \begin{subfigure}[b]{\textwidth}
\setlength{\unitlength}{1\textwidth}
       \begin{picture}(1,0.03)

       \put (0.85,0.0005) {\textbf{\color{red} Q1}}
       \put (0.67,0.0005) {\textbf{\color{red} Q2A}}
        \put (0.48,0.0005) {\textbf{\color{red} Q2B}}
        \put (0.3,0.0005) {\textbf{\color{red} Q3}}      
        \put (0.12,0.0005) {\textbf{\color{red} D1}} 
       \end{picture}
      \centering
       \includegraphics[trim=0.0cm 0.0cm 0.0cm 0.0cm,width=0.18\textwidth]{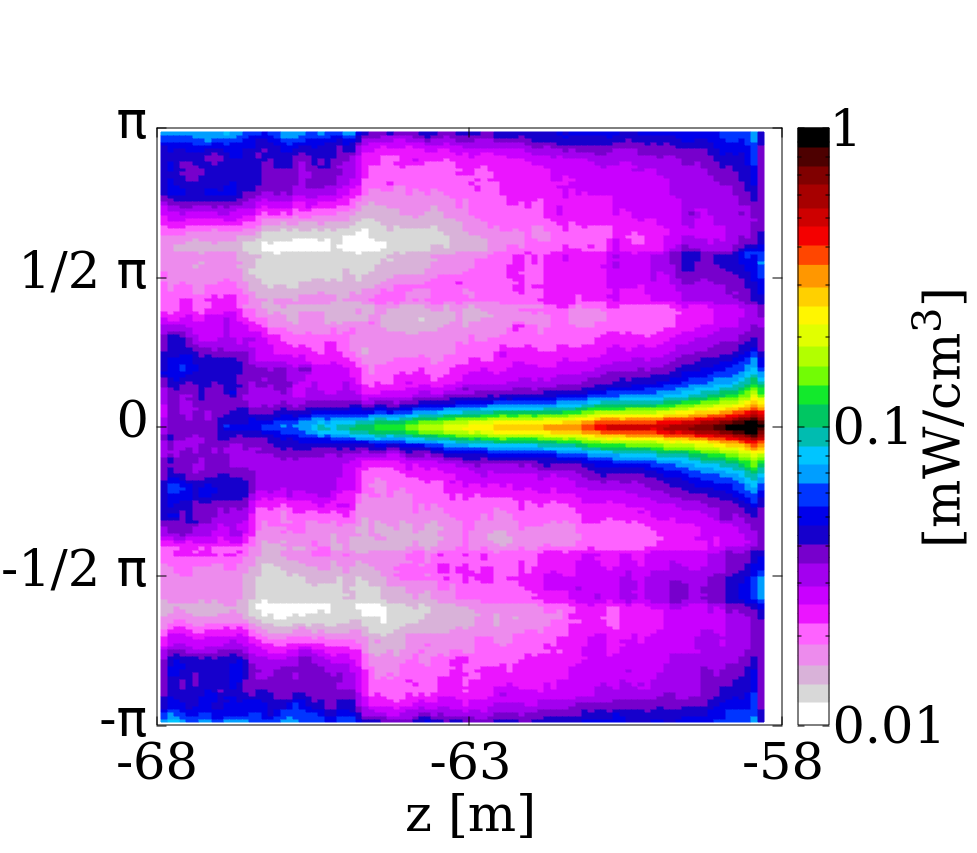}
       \includegraphics[trim=0.0cm 0.0cm 0.0cm
       0.0cm,width=0.18\textwidth]{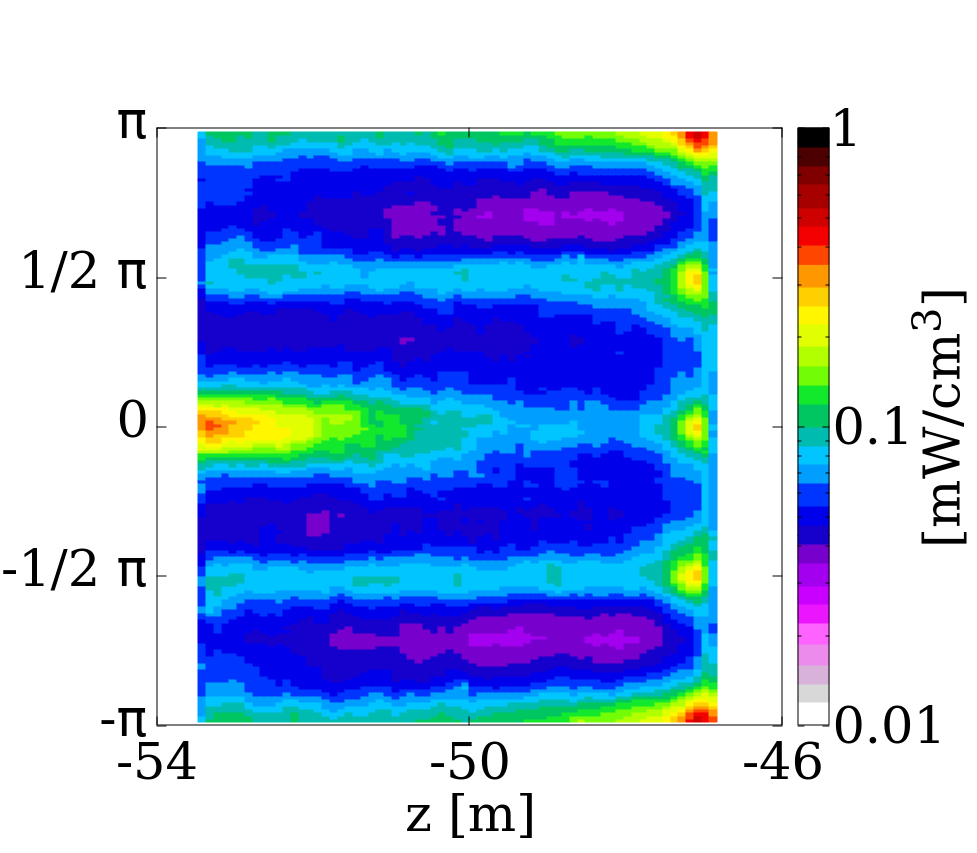}
       \includegraphics[trim=0.0cm 0.0cm 0.0cm 0.0cm,width=0.18\textwidth]{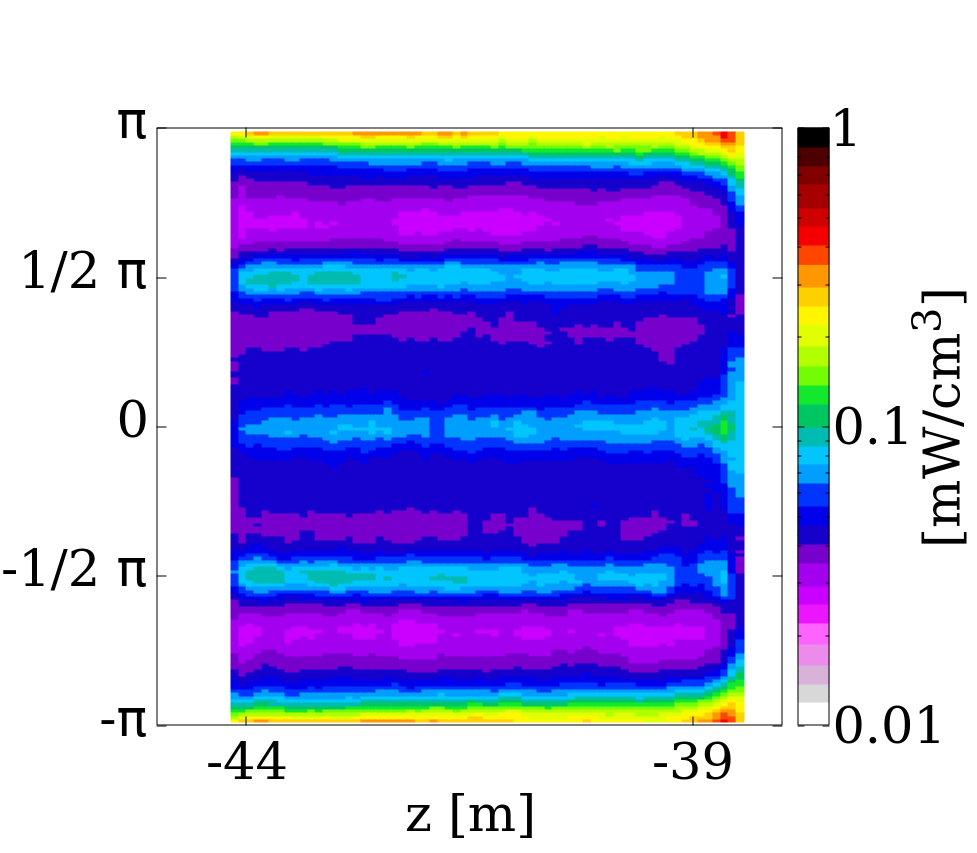}
       \includegraphics[trim=0.0cm 0.0cm 0.0cm 0.0cm,width=0.18\textwidth]{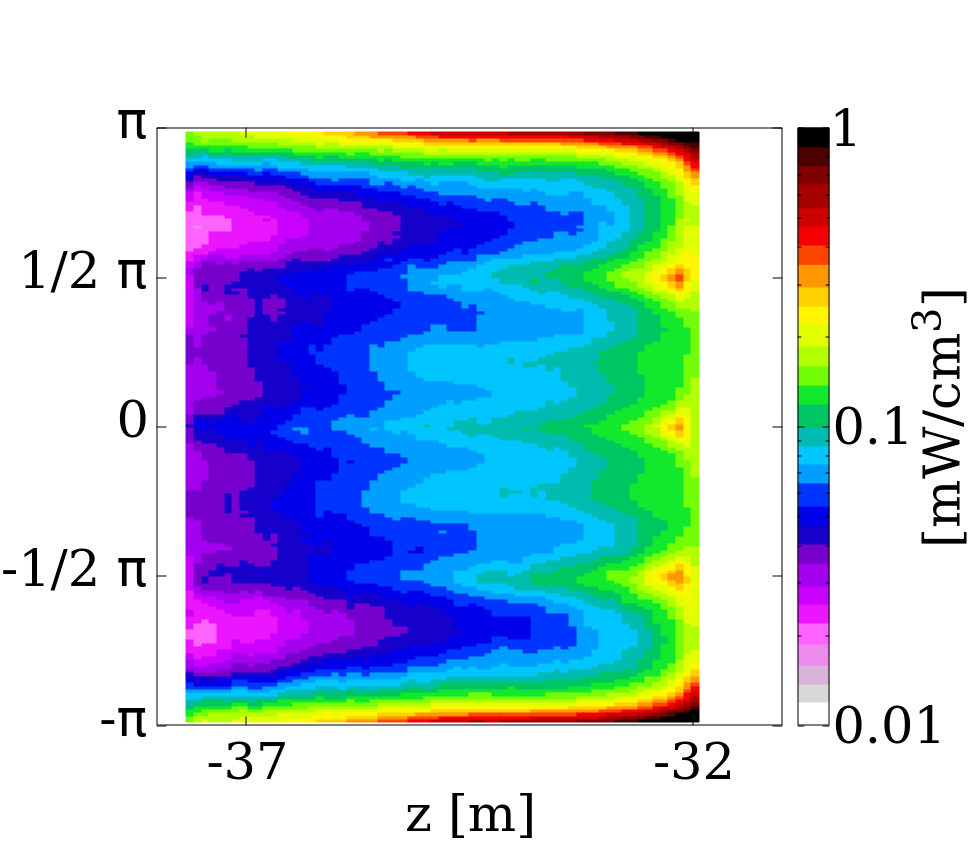}
       \includegraphics[trim=0.0cm 0.0cm 0.0cm 0.0cm,width=0.18\textwidth]{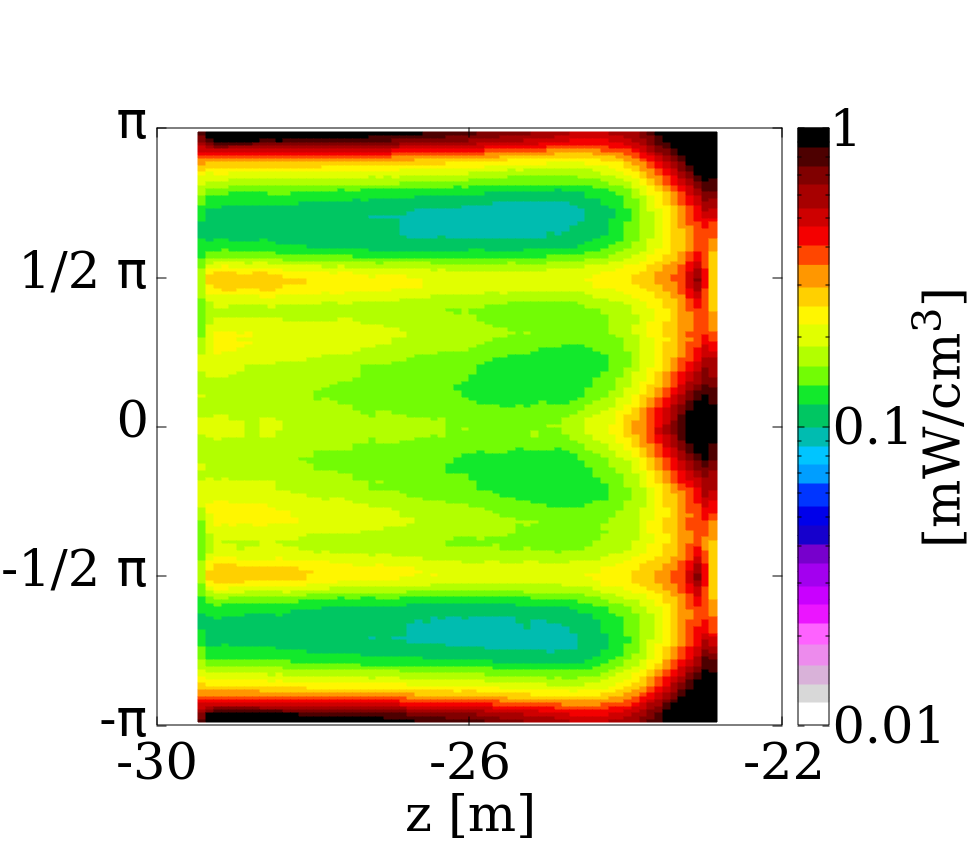}
       \includegraphics[trim=0.0cm 0.0cm 0.0cm 0.0cm,width=0.18\textwidth]{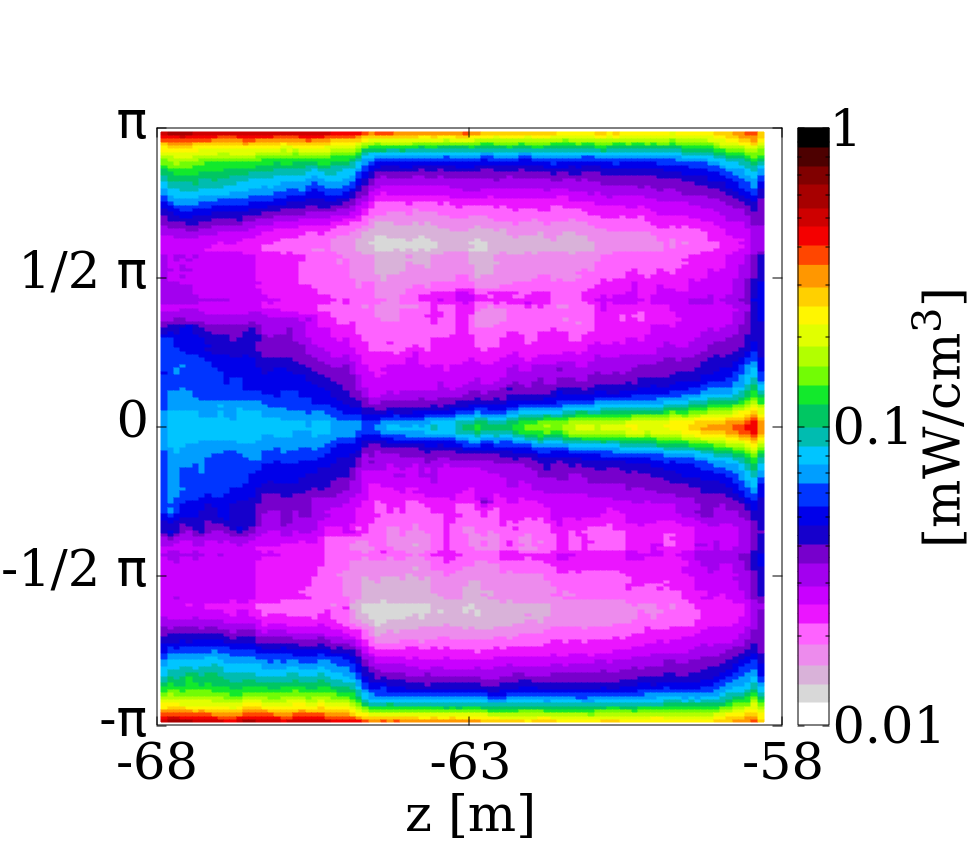}
       \includegraphics[trim=0.0cm 0.0cm 0.0cm
       0.0cm,width=0.18\textwidth]{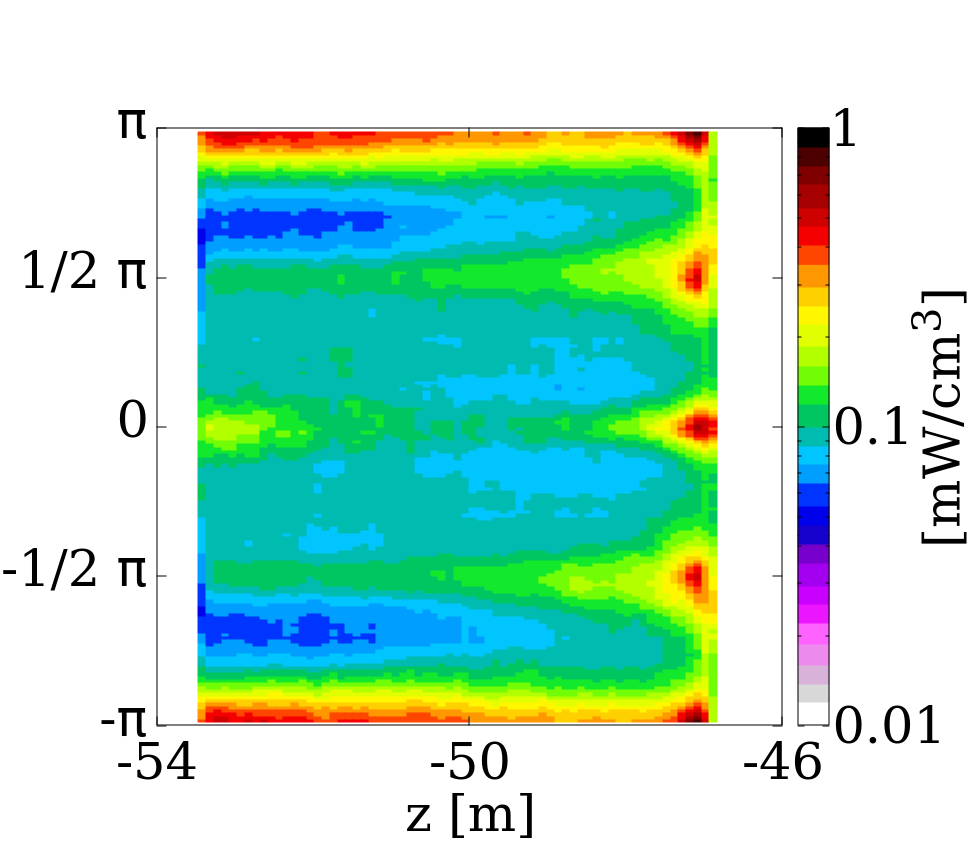}
       \includegraphics[trim=0.0cm 0.0cm 0.0cm 0.0cm,width=0.18\textwidth]{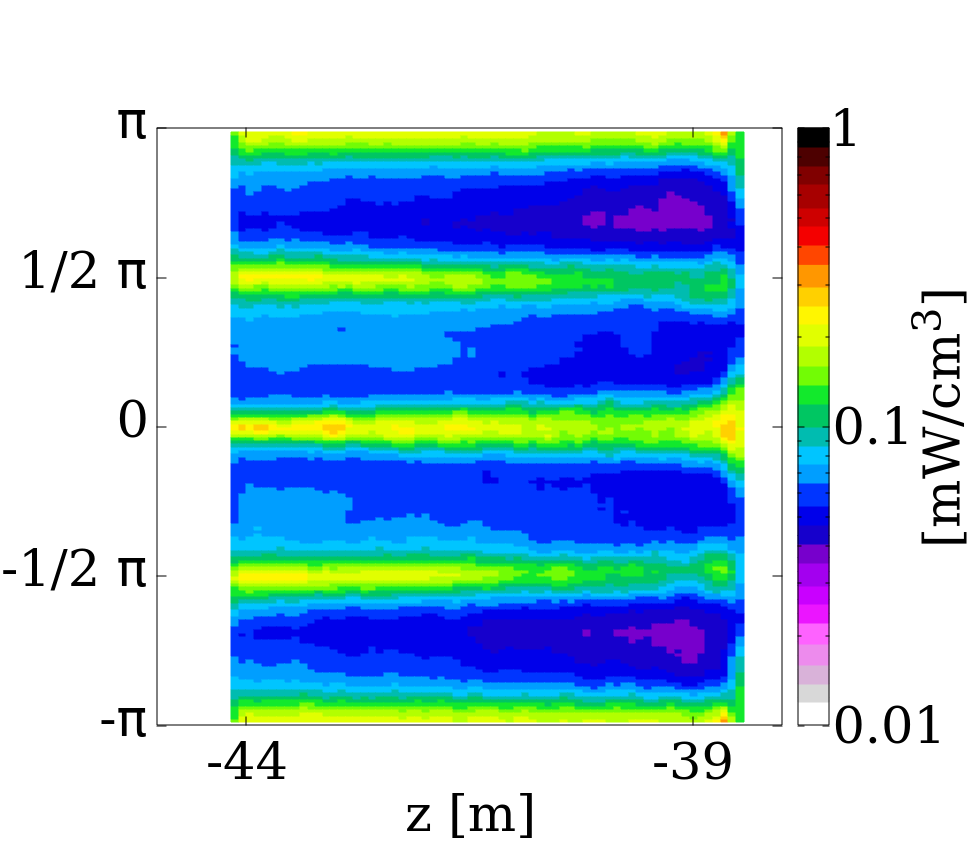}
       \includegraphics[trim=0.0cm 0.0cm 0.0cm 0.0cm,width=0.18\textwidth]{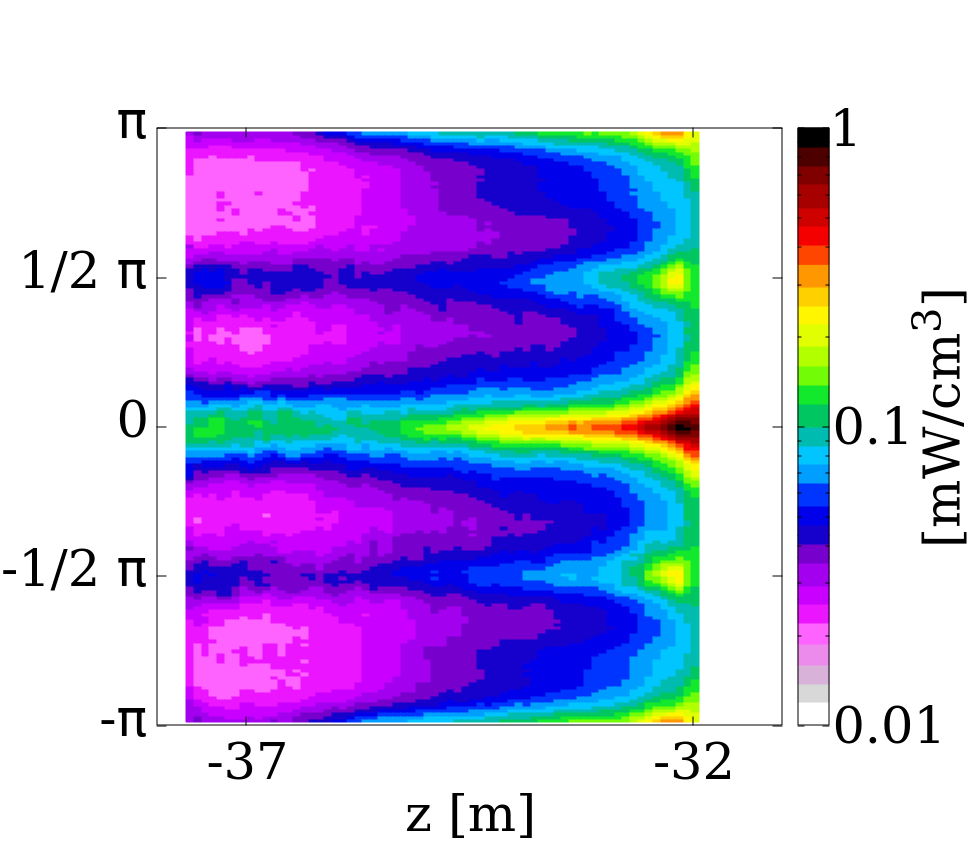}
       \includegraphics[trim=0.0cm 0.0cm 0.0cm 0.0cm,width=0.18\textwidth]{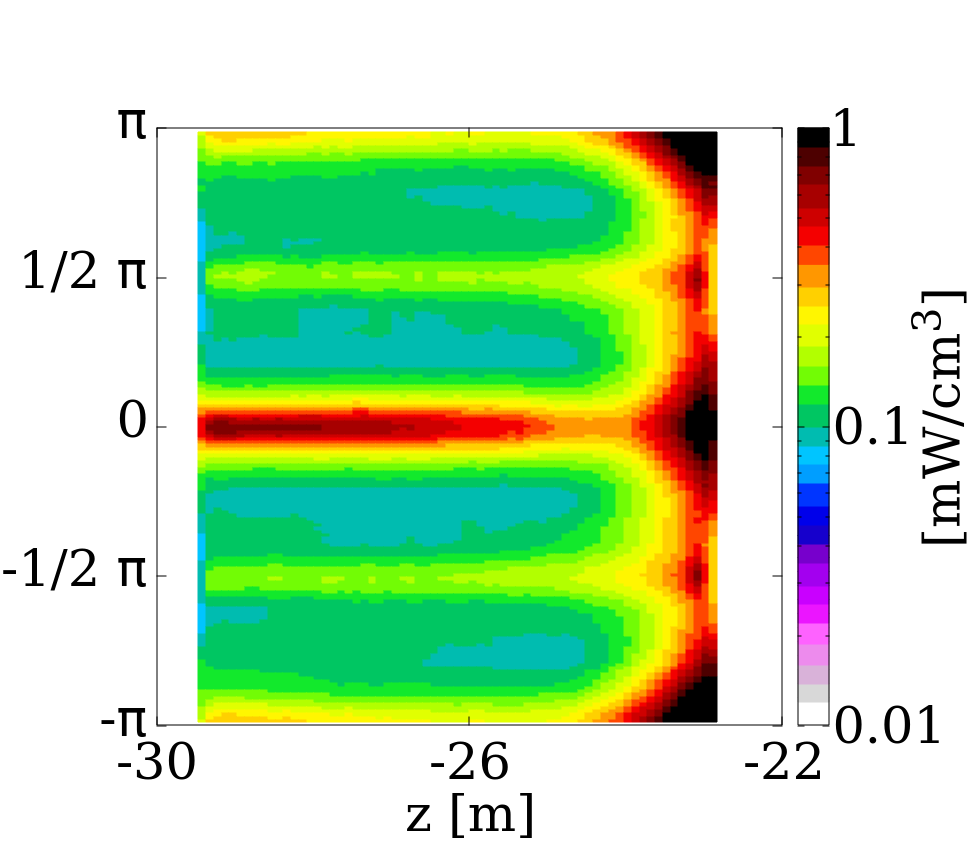}
       \begin{picture}(1,0.01)
       \put(0.74,0.01){\vector(-1,0){0.08}}
       \put(0.75,0.005){Outgoing beam (B2)}
       \end{picture}
        \caption{The same as Fig.~\ref{fig:vert} with an external half crossing angle of $250\,\mathrm{\mu}$rad in the horizontal plane.}
        \label{fig:hor}
\end{figure*} 
The spectra for neutrons and photons travelling inside the vacuum chamber are displayed in Figs~\ref{fig:neutron}-\ref{fig:photon}. They are not affected by magnetic fields and propagate in a straight line until they meet an aperture restriction. As shown in Fig.~\ref{fig:neutron}, TeV neutrons travel undisturbed beyond the separation dipole, because of their highly forward angular distribution, and are later intercepted by the TANB absorber between the two separate beam tubes.
The photon spectra in Fig.~\ref{fig:photon} have a broad peak around few hundred~GeV. As indicated by the difference between the pink and red curves, a sizeable quantity of photons is absorbed by the Q1, due to the absence of the TAS. The aperture of the following quadrupoles is larger and so puts the Q2A in the geometrical shadow of the Q1, which ends in the Q2B. Thereby, the latter is also subject to the photon impact, as visualized by the difference between the green and the blue curves. 

Considering the spacial evolution of the debris along the triplet and the D1, we find that a crucial role is played by the crossing scheme, coupled with the triplet magnetic configuration. The understanding of this point is also important in order to work out how possible combinations of different schemes can minimize the coil insulator degradation due to radiation and so increase the magnet lifetime.
The power density distribution in the inner layer of the superconducting coils, depending on the polarity of the LHCb spectrometer, is shown in Fig.~\ref{fig:vert} and Fig.~\ref{fig:hor} for external crossing in the vertical and horizontal plane, respectively.  
The debris leaves the IP around the direction of the outgoing beam, as determined by the actual crossing angle. This results from the superposition of the spectrometer bump on the horizontal plane and the external crossing enabled by the orbit corrector magnets. If the external crossing is horizontal (as in Fig.~\ref{fig:hor}), the crossing angle in IP8 sits on the horizontal plane. If the external crossing is vertical (as in Fig.~\ref{fig:vert}), the crossing plane in IP8 is skew. In other words, the latter is never vertical inside the LHCb experiment. 
The presence of the warm compensator magnets and the LHCb dipole, acting only in the horizontal plane, makes the charged debris to be intercepted mainly on the horizontal plane, for either external crossing option. The peak power density in the Q1 and first half of Q2 lies on the inside of the ring ($\phi=\pm\pi$) for LHCb upward polarity and on the outside ($\phi=0$) for LHCb downward polarity. As an effect of the triplet field, a reversal of the peak position takes place later, as clearly visible at the non-IP extremity of Q3. 
In case of external crossing on the vertical plane, another lower peak is found in Q1 at $\phi=\pi/2$. Moreover, for LHCb downward polarity (bottom plots of Fig.~\ref{fig:vert}) the positively charged debris, which is concentrated around $\phi=\pm\pi$ at the D1 entrance, is further pushed by the separation dipole field towards the inside of the ring, and this significantly amplifies its impact on the D1 coils, as made apparent afterward in Fig.~\ref{fig:Run3Power}. 
\section{Run~2}
\label{sec:run2}
\subsection{Beam loss monitor measurements}
The BLM system is an essential part of the machine protection architecture to ensure safe LHC operation~\cite{HOLZER20122055}. The beam losses are monitored in real time through the dose values collected in the BLMs. These are cylindrical ionization chambers featuring parallel aluminum electrode plates and filled with nitrogen gas. More than 3600 BLMs are placed around the LHC in selected locations. The signals are converted to dose rate in $\mathrm{Gy}\,\mathrm{s^{-1}}$. The front-end electronics provides 12 output signals (running sums 'RS') corresponding to integration periods from $40\,\mu\mathrm{s}$ to $84\,\mathrm{s}$. Beam losses along the accelerator may induce BLM dose rate values exceeding pre-defined threshold and so trigger a beam dump request, which is meant to prevent cold magnet quenching or equipment (e.g., collimator) damage.
The BLM thresholds are set depending on beam energy and loss duration, in relation to possible hazardous losses originated by the circulating beams. For this study, the BLM signals with the maximum integration time were post-processed subtracting the noise floor from the measured signals in order to accurately measure low doses. This technique has been already used in the previous BLM benchmark studies with FLUKA~\cite{PhysRevAccelBeams.22.071003}.  
\subsection{Simulation benchmarking}
\begin{figure}[!ht]
      \includegraphics[width=1\linewidth]{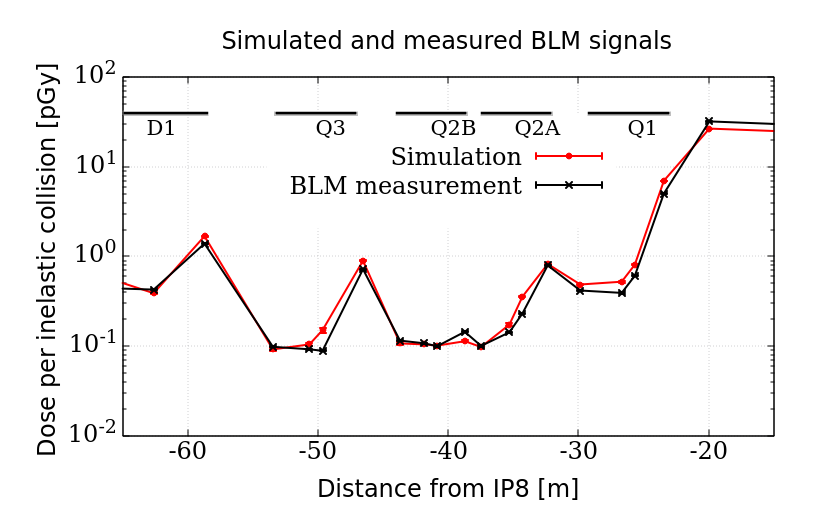}
      \caption{Simulated (in red) and measured (in black) BLM signals along (from right-to-left) the four quadrupoles Q1-Q2A-Q2B-Q3 and the D1 separation dipole located on the left side of IP8. All dose values are given per inelastic nuclear interaction generated by $6.5\,\mathrm{TeV}$ proton beams colliding in IP8 on the horizontal plane with a half crossing angle of $-395\,\mathrm{\mu}$rad (pointing inside the ring). The experimental data are the result of averaging signals of several fills recorded from 14/08/2018 to 11/10/2018.}
      \label{fig:Q1-D1}
\end{figure}
The simulation has been benchmarked against BLM signals measured during proton physics fills in Run~2. %https://abpdata.web.cern.ch/abpdata/lhc_optics_web/www/opt2018/
The simulated configuration corresponds to the Run~2 layout with the 2018 $6.5\,\mathrm{TeV}$ optics. An external horizontal crossing of $-250\,\mathrm{\mu}$rad (with beams pointing inside the ring) was assumed to be coupled to the LHCb spectrometer downward polarity. The latter implies that the incoming beam is further deflected by $-145\,\mathrm{\mu}$rad. 

Fig.~\ref{fig:Q1-D1} presents the comparison between measurements and simulation predictions in terms of dose per $13\,\mathrm{TeV}$ center-of-mass inelastic collision. A previous benchmark study focused on the right side triplet of both LHCb and ATLAS, with regard to Run~1 operation with $4~\mathrm{TeV}$ proton beams~\cite{Lechner:2674116}. The authors found that simulated signals were on average $20\%$ and $50\%$ higher than data in IR1 and IR8, respectively. The larger overestimation in IR8 was tentatively attributed to secondary particles generated upstream of the Q1 and reaching the BLMs by travelling outside the magnets. The artificial suppression of their contribution in the simulation led to values lower than measurements. In reality, some of these particles are intercepted by external material not included in the simulation model. In this work, special care was devoted to refine the FLUKA geometry, especially on the left side, where the aperture of the shielding wall between the two warm dipoles (MBWLH and MBLWS) turned out to play a crucial role to our benchmarking purposes. Assuming a square hole of 12 cm edge in the aforementioned shielding around the beam pipe, the resulting agreement is within $20\%$, compatibly with value uncertainties.

\section{Run~3: predictions for the upcoming high luminosity era}
\label{sec:run3}
An important outcome of this study is the review of the impact of the Upgrade I of LHCb on the accelerator elements in IR8.
During 2022, the first year of Run~3, the external crossing at IP8 will be in the horizontal plane, with an integrated luminosity forecast of $4\,\mathrm{fb}^{-1}$ at $6.8\,\mathrm{TeV}$ proton beam energy. Then, from 2023 to 2025, the external crossing shall be switched to the vertical plane, aiming to record additional $24\,\mathrm{fb}^{-1}$. So the latter is the more relevant scenario, we simulated with a $+200\,\mathrm{\mu}$rad half angle, implying a vertical momentum component towards the top. This was applied to $7\,\mathrm{TeV}$\footnote{At the time of these studies, the actual Run~3 beam energy was not yet defined.} beams and combined with:
   \begin{itemize}
     \item a $+135\,\mathrm{\mu}$rad kick for upward polarity of the LHCb spectrometer, giving in IP8 a half crossing angle of $\simeq\!240\,\mathrm{\mu}$rad on a skew half plane oriented at $\phi=56^{\circ}$ (see the caption of Fig.~\ref{fig:vert} for the azimuthal angle definition); 
     \item a $-135\,\mathrm{\mu}$rad kick for downward polarity of the LHCb spectrometer, giving in IP8 the same half crossing angle of $\simeq\!240\,\mathrm{\mu}$rad on a skew half plane oriented at $\phi=124^{\circ}$.
   \end{itemize}
\begin{figure}[!ht]
      \centering
       \includegraphics[trim=1.0cm 1.0cm 2.0cm 3.5cm, clip=true,width=0.95\linewidth]{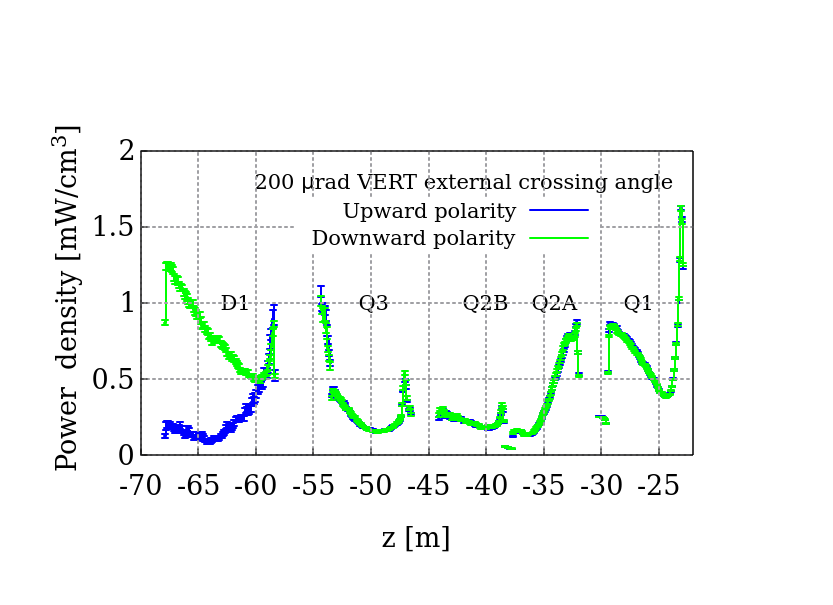}
      \caption{Longitudinal profile of peak power density in the superconducting coils along the triplet and the D1 (from right to left) on the left side of IP8 (at z=0). Values are averaged over the cable radial thickness and normalized to the indicated instantaneous luminosity (representing the Run~3 target). The azimuthal resolution is of 2$^{\circ}$. External vertical crossing has been simulated for $\sqrt{s}=14\,\mathrm{TeV}$ in combination with either upward (blue points) and downward (green points) polarity of the LHCb spectrometer.}
      \label{fig:Run3Power}
\end{figure}
\begin{figure}[!ht]
      \centering
       \includegraphics[trim=1.0cm 1.0cm 2.0cm 3.5cm,clip,width=0.95\linewidth]{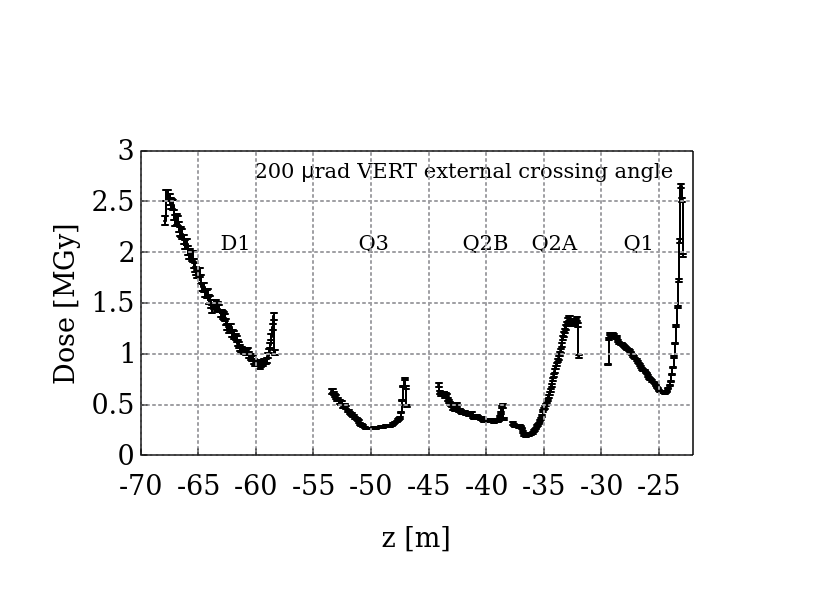}
      \caption{Longitudinal profile of peak dose in the superconducting coils. The azimuthal and radial resolution is of 2$^{\circ}$ and 3 mm, respectively. Values refer to external vertical crossing and an integrated luminosity of $24\,\mathrm{fb}^{-1}$, half of which collected with either polarity of the LHCb spectrometer at $\sqrt{s}=13.6\,\mathrm{TeV}$.}
      \label{fig:Run3Dose}
\end{figure}
\subsection{The triplet and separation dipole}

\begin{figure*}[!ht]
      \centering
      \includegraphics[trim=3.0cm 0.5cm 1.0cm 
      1.8cm,clip,width=0.32\linewidth]{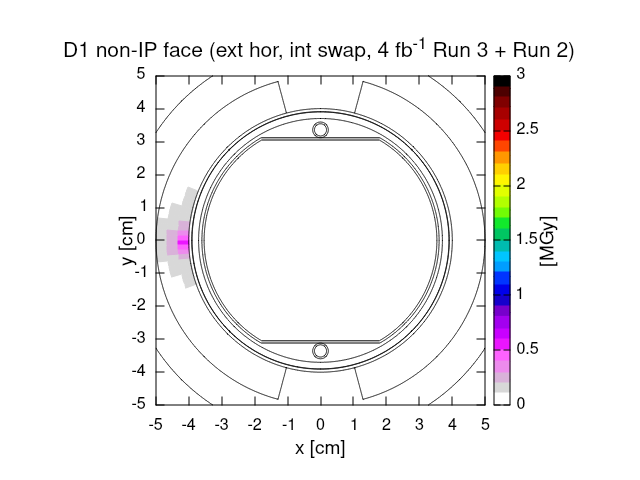}
%    \hfill 
       \includegraphics[trim=3.0cm 0.5cm 1.0cm 1.8cm,clip,width=0.32\linewidth]{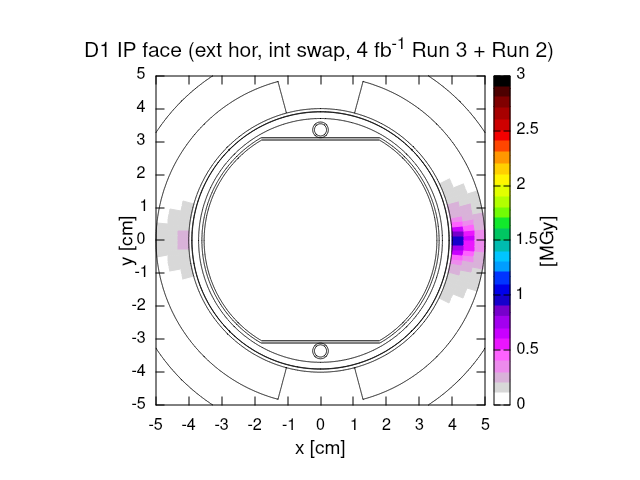}
%    \hfill 
       \includegraphics[trim=3.0cm 0.5cm 1.0cm 1.8cm,clip,width=0.32\linewidth]{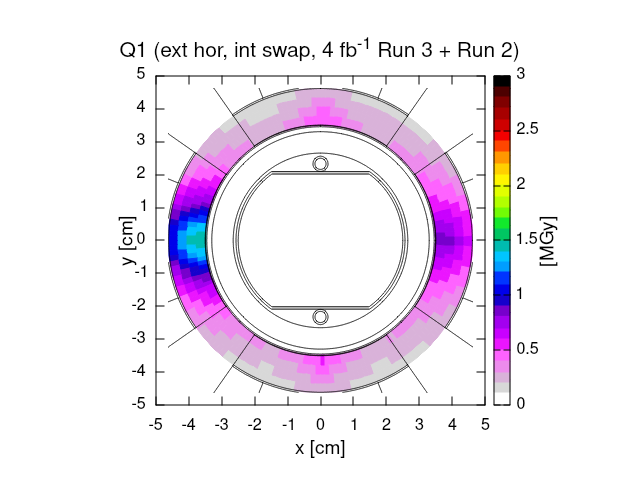}
       \caption{Transverse dose distribution on the D1 non-IP face (left), D1 IP face (center) and Q1 IP face (right) for external crossing in the horizontal plane. The contribution of both 2022 operation ($4\,\mathrm{fb}^{-1}$ at $\sqrt{s}=13.6\,\mathrm{TeV}$) and Run~2 ($6.6\,\mathrm{fb}^{-1}$ at $\sqrt{s}=13\,\mathrm{TeV}$ from 2015 to 2018) is included, assuming that half of the respective integrated luminosity is produced with either LHCb polarity.}
       \label{fig:hor_dose}
\end{figure*}

The first aspect to assess is the operational margin with respect to the quench limit. To do so, one has to evaluate the maximum power density deposited in the superconducting coils, which in steady state conditions is usually calculated by averaging over the cable transverse area.  
\begin{figure*}[ht!]
      \begin{center}
%      \hfill
      \includegraphics[trim=3.0cm 0.5cm 1.0cm 1.8cm,clip,width=0.32\linewidth]{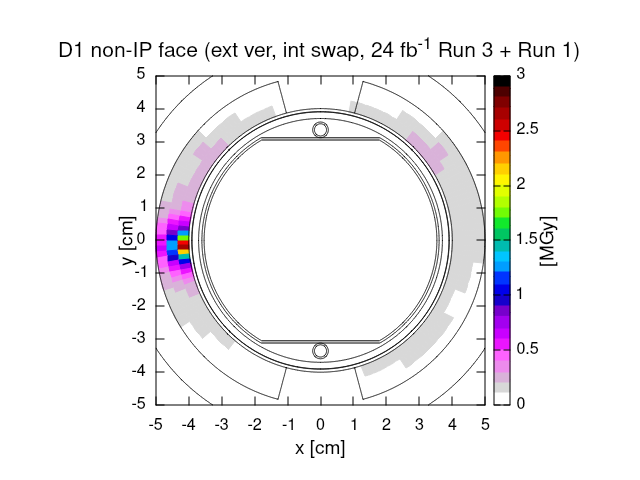}
       \includegraphics[trim=3.0cm 0.5cm 1.0cm 1.8cm,clip,width=0.32\linewidth]{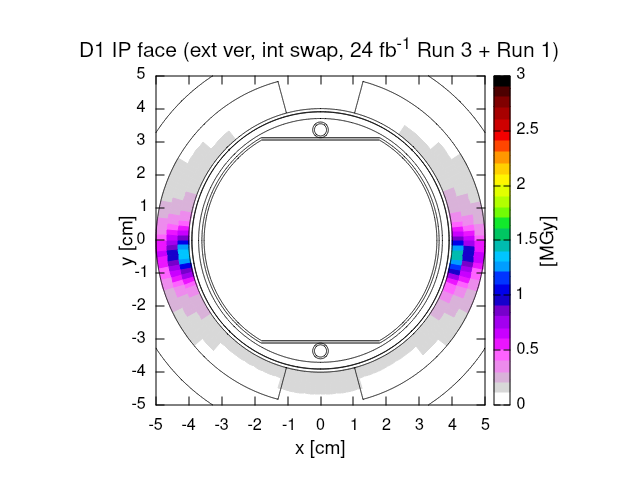}
 %\hfill  
       \includegraphics[trim=3.0cm 0.5cm 1.0cm 1.8cm,clip,width=0.32\linewidth]{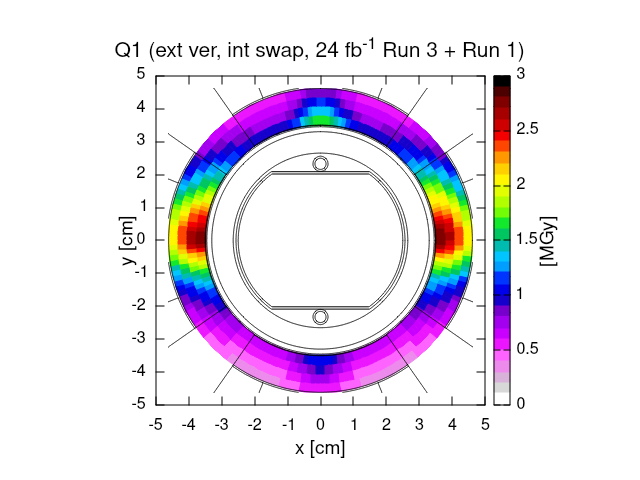}
       \caption{Transverse dose distribution on the D1 non-IP face (left), D1 IP face (center) and Q1 IP face (right) for external crossing in the vertical plane (with $+200\,\mathrm{\mu}$rad half angle). The contribution of both 2023 to 2025 operation ($24\,\mathrm{fb}^{-1}$ at $\sqrt{s}=13.6\,\mathrm{TeV}$) and Run~1 ($3\,\mathrm{fb}^{-1}$ at $\sqrt{s}=7-8\,\mathrm{TeV}$) is included, assuming that half of the respective integrated luminosity is produced with either LHCb polarity.}
       \label{fig:vert_dose}
       \end{center}
\end{figure*} 
%trim={<left> <lower> <right> <upper>}
On the other hand, Fig.~\ref{fig:hor_dose} reports the picture for external crossing in the horizontal plane, as adopted during Run~2 and planned for 2022.
Fig.~\ref{fig:Run3Power} shows the peak power density profile along the triplet and the D1 dipole on the left side of IP8, for the two polarities of the LHCb spectrometer. Both configurations display an absolute maximum at the IP side of Q1, due to the absence of the TAS. For the Run~3 instantaneous luminosity of 2 $10^{33}\,\mathrm{cm}^{−2}\,\mathrm{s}^{−1}$, its value is less than one half of the design limit. In fact, the quench limit for the triplet was estimated to be $1.6\,\mathrm{mW/g}$, namely $13\,\mathrm{mW/cm^{3}}$, and a safety factor of three was taken in defining the design limit at $4.3\,\mathrm{mW/cm^{3}}$~\cite{Mokhov:613167}. Nevertheless, our result confirms that a Q1 protection strategy is required in view of the Upgrade II of LHCb, as the instantaneous luminosity further increases by a factor of 7.5, driving the Q1 peak very close to the quench limit and well beyond the values already reached in the IR1 and IR5 triplets with a luminosity doubling the ATLAS and CMS design value. Looking at the debris behaviour shown in Fig.~\ref{fig:vert}, one can see that the peaks are located in the horizontal plane, on opposite sides for the two spectrometer polarities. The peak profile is the same up to the D1 IP face, where the trend changes. There, in the configuration corresponding to the downward polarity, the dominant component of the debris, positively charged, is concentrated toward the inside of the ring, as a result of its passage through the magnetic field of the triplet. As the separation dipole deflects positive particles into the ring, debris losses rise along the magnet up to reaching $1.25\,\mathrm{mW/cm^{3}}$ at the non-IP end of the D1. Like for the Q1, this is not alarming for Run~3, but should be addressed in view of the Upgrade~II.

The other important aspect to be studied is the material degradation due to the radiation exposure. In particular, the long term deterioration of the coil insulator, as a function of the radiation dose accumulated with the integrated luminosity, can jeopardize the magnet functionality and so determines its lifetime. Fig.~\ref{fig:Run3Dose} shows the peak dose profile that is expected to be produced during the last three years of Run~3, assuming external crossing in the vertical plane and an equal sharing of the integrated luminosity target between the two LHCb polarity configurations.
The 2D dose map for the three highest peaks is displayed in Fig.~\ref{fig:vert_dose}, after adding the Run~1 contribution.
 
In total, the maximum dose predicted on the IP face of Q1 by the end of Run~3 in 2025 is $4.5\,\mathrm{MGy}$ on the left of IP8, rising to about $6\,\mathrm{MGy}$ on the right of IP8, where the triplet is more exposed. As far as D1 is concerned, the maximum dose is $3\,\mathrm{MGy}$ on the non-IP face and $2.5\,\mathrm{MGy}$ on the IP face. These numbers are safely below the damage limit of 30 MGy that is known to apply to Q1-Q3~\cite{Tavlet}. Nonetheless, their increase for the ultimate LHCb Upgrade II target of $400\,\mathrm{fb}^{-1}$ requires a dedicated analysis, specifically on the corrector magnets embedded in the triplet, whose multi-wire cable insulation may start to degrade already over a dose range not exceeding $10\,\mathrm{MGy}$.

\subsection{The warm compensators}
As discussed in Section~\ref{sec:model}, the warm magnets compensating for the LHC spectrometer kick are the closest elements to IP8. The one most impacted by the collision debris is the short compensator on the right side, absorbing $13.5\,\mathrm{W}$ at the Run~3 instantaneous luminosity. Despite its larger mass, the long compensator MBXWH, well shielded by a concrete wall which seals off the experimental cavern on the left side, gets only $11\,\mathrm{W}$. In fact, thanks to its proximity to IP8, it is also missed by the most energetic particles that travel at too low angles with respect to the longitudinal axis. 
\begin{figure}[!h]
      \centering
       \includegraphics[trim=1.0cm 1.0cm 1.0cm 1.0cm, clip=true,width=0.95\linewidth]{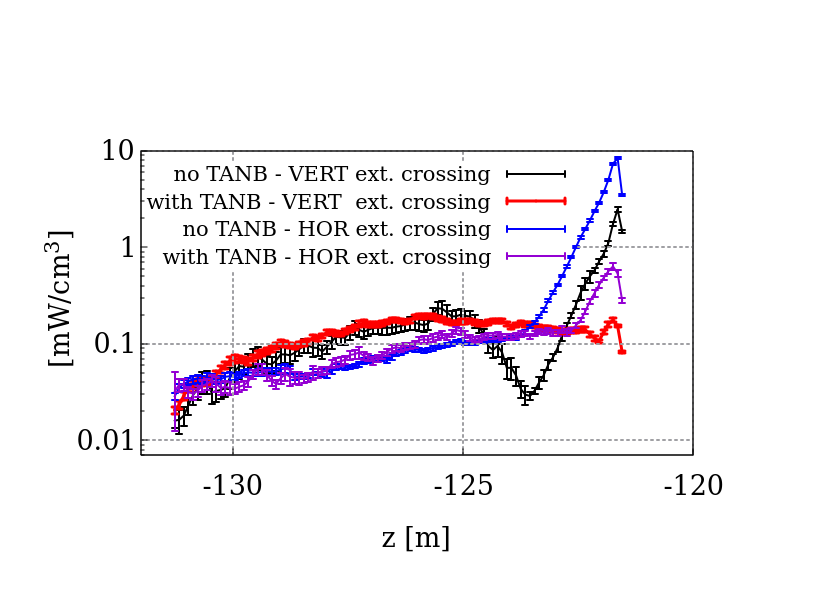}
      \caption{Longitudinal profile of peak power density in the D2 superconducting coils on the left side of IP8 (at z=0). Values are averaged over the cable radial thickness and normalized to the indicated instantaneous luminosity (representing the Run~3 target). The azimuthal resolution is of 2$^{\circ}$. Four cases have been simulated for $\sqrt{s}=14\,\mathrm{TeV}$ and downward LHCb polarity: external crossing in the horizontal or vertical plane, with or without the TANB.}% BETTER LOG Y SCALE?
      \label{fig:Run3TANb}
\end{figure} 
A weak point is represented by the coils on the magnet IP face, especially for the short compensator on the right side. The peak dose reached just above the vacuum pipe (assuming external vertical crossing with $+200\,\mathrm{\mu}$rad half angle) is predicted to surpass $10\,\mathrm{MGy}$ by the end of Run~3. These findings suggest implementing before the end of Run~3 a suitable tungsten piece acting as coil protection, in order to reduce the maximum dose by a factor of few, also in view of the further luminosity increase later envisaged. A minor gain may come from the polarity inversion of the external crossing angle.

\subsection{The recombination dipole}
The first HL-LHC object installed in the machine during LS2 was the TANB, as earlier mentioned in Section~\ref{sec:model}. It was designed to shield the D2 recombination dipole on both sides of IP8 from forward high-energy neutral particles produced by proton--proton collisions.
As a result, the total power absorbed by the D2 cold mass at $2\cdot10^{33}\,\mathrm{cm}^{−2}\,\mathrm{s}^{−1}$ decreases from $30\mathrm{W}$~\cite{Esposito:1636136} to $6\mathrm{W}$ (less on the right side of IP8, because of the additional protection provided by the TCDDM mask). Moreover, Fig.~\ref{fig:Run3TANb} shows the TANB effect on the peak power density in the D2 superconducting coils. The maximum on the IP face is reduced by more than a factor 10 for both external crossing schemes. This translates into a maximum dose to the coil insulator lower than $1\,\mathrm{MGy}$ by the end of Run~3 (2025), after additional $28\,\mathrm{fb}^{-1}$ at $\sqrt{s}=13.6\,\mathrm{TeV}$. Thanks to its proximity to the recombination dipole, the TANB may also fulfil its protection functions in the Upgrade II scenario, as current studies confirm.
\\
\section{Conclusions and outlook}
\label{sec:conclusions}
In this paper we have evaluated radiation levels induced by proton beam collisions in the LHCb detector. The FLUKA model of IR8 has been improved and further validated by comparing the BLM dose values measured in 2018 physics fills with the simulation predictions in the region of the triplet and separation dipole.
The obtained good agreement corroborates the FLUKA model reliability for addressing the challenges raised by the future luminosity increase. In particular, we used then model to review the implications of running LHCb at $2\cdot10^{33}\,\mathrm{cm}^{−2}\,\mathrm{s}^{−1}$, as planned in the upcoming Run~3 of the LHC. 
The collision debris from proton operation at that instantaneous luminosity, about 5 times higher than in Run~2, is predicted not to pose any threat with respect to quench limits and cryo-capacity. On the other hand,
after accumulating an additional $28\,\mathrm{fb}^{-1}$ by the end of 2025, a maximum dose of about $6\,\mathrm{MGy}$ is expected to be reached in the Q1 coils on the right of IP8. The most exposed high order correctors embedded in the triplet would get $2-3\,\mathrm{MGy}$. In parallel, the front coils of the short warm compensators would reach $10-12\,\mathrm{MGy}$. The installation of tungsten shields, as already implemented in the IR7 collimation insertion, appear to be a viable mitigation solution. The recombination dipole benefits from the TANB effective protection.

The investigation of the Upgrade II scenario is currently ongoing, aiming to indicate effective solutions that allow accelerator operation at $1.5\cdot10^{34}\,\mathrm{cm}^{−2}\,\mathrm{s}^{−1}$ and for $400\,\mathrm{fb}^{-1}$ in LHCb. This calls for the comprehensive study of several measures. Some of them naturally follow the conclusions of this work, such as the shielding of the short compensators, the Q1 quadrupoles and D1 separation dipoles, and the TANB cooling.
Other required work is the construction of a new wall in the UX85 cavern~\cite{2424228} and the possible integration of physics debris collimators (TCL) in the machine around LHCb.

\begin{acknowledgments}
Research supported by the HL-LHC project. We thank Riccardo De Maria and Stephane Fartoukh for providing us with optics input as well as François Butin, Francisco Sanchez Galan and Maud Wehrle for making available relevant information as well as the technical drawings that were crucial for the geometry description development. We also wish to acknowledge the support of the LHCb Collaboration, who kindly shared the FLUKA model of the LHCb detector.
\end{acknowledgments}

%\nocite{*}
%\FloatBarrier
\bibliography{apssamp}% Produces the bibliography via BibTeX.

\end{document}